\documentclass[11pt]{article}

\pdfoutput=1 
\usepackage{jheppub} 

\usepackage{graphicx}
\usepackage{amsmath, amssymb}

\usepackage{tikz}

\usepackage{verbatim}
\usepackage{caption}
\usepackage{subcaption}

\newcommand{\be}{\begin{eqnarray}}
\newcommand{\ee}{\end{eqnarray}}
\newcommand{\bea}{\begin{eqnarray}}
\newcommand{\eea}{\end{eqnarray}}

\newcommand{\bn}{\begin{enumerate}}
\newcommand{\en}{\end{enumerate}}

\def\CN{{\cal N}}

\def\CS{{\cal S}}

\def\Tr{\mathop{\mathrm{Tr}}\nolimits}
\def\diag{\mathop{\mathrm{diag}}}
\def\SU{\mathrm{SU}}
\def\SO{\mathrm{SO}}
\def\Spin{\mathrm{Spin}}
\def\USp{\mathrm{USp}}
\def\U{\mathrm{U}}
\def\bC{\mathbb{C}}
\def\bZ{\mathbb{Z}}

\def\half{\frac{1}{2}}

\def\Nequals#1{$\mathcal{N}{=}#1$}
\def\fsu{\mathfrak{su}}
\def\fso{\mathfrak{so}}
\def\gcd{\mathop{\mathrm{gcd}}}
\def\fu{\mathfrak{u}}

\title{4d \Nequals1 from 6d \Nequals{(1,0)} on a torus with fluxes}

\preprint{IPMU17-0013\;\;\;\;\; Imperial/TP/17/AH/02}

\author[a,b]{Ibrahima Bah,}
\author[c]{Amihay Hanany,}
\author[d]{Kazunobu Maruyoshi,}
\author[e]{Shlomo S. Razamat,}
\author[f]{\\ Yuji Tachikawa,}
\author[f]{and Gabi Zafrir}

\affiliation[a]{Department of Physics, University of California, San Diego, La Jolla, CA 92093 USA}
\affiliation[b]{Department of Physics and Astronomy, Johns Hopkins University,\\ 3400 North Charles Street, Baltimore, MD 21218, USA}
\affiliation[c]{Imperial College London, Blackett Laboratory, Prince Concert Road,\\ South Kensington, London, SW7 2AZ, UK}
\affiliation[d]{Faculty of Science and Technology, Seikei University,\\
3-3-1 Kichijoji-Kitamachi, Musashino-shi, Tokyo, 180-8633, Japan}
\affiliation[e]{Department of Physics, Technion, Haifa, 32000, Israel}
\affiliation[f]{IPMU,  University of Tokyo,  Kashiwa, Chiba 277-8583, Japan}
\emailAdd{Iboubah@jhu.edu}
\emailAdd{a.hanany@imperial.ac.uk}
\emailAdd{maruyoshi@st.seikei.ac.jp}
\emailAdd{razamat@physics.technion.ac.il}
\emailAdd{gabi.zafrir@ipmu.jp}
\emailAdd{yuji.tachikawa@ipmu.jp}

\abstract
{Compactifying \Nequals{(1,0)} theories on a torus, with additional fluxes for global symmetries, we obtain \Nequals1 supersymmetric theories in four dimensions.
It is shown that for many choices of flux these models are toric quiver gauge theories with singlet fields. In particular we compare the anomalies deduced from the description of the six-dimensional theory
and the anomalies of the quiver gauge theories. We also give predictions for anomalies of four-dimensional theories corresponding to general compactifications of M5-branes probing $\bC^2/\bZ_k$ singularities.}

\begin{document} 
\maketitle
\flushbottom



\

\

\

\section{Introduction}  Field theories in low dimensions can often be realized through compactifications of higher dimensional models. This point of view clarifies some of the well known properties of quantum field theories, and also predicts new properties and even new models; for example, the appearance of theories which do not have a known semiclassical limit. Such models are ubiquitous in compactifications of six-dimensional supersymmetric theories to four dimensions \cite{Gaiotto:2009we}.

\

In this paper we mainly study some of the simpler compactifications. We consider $N$ M5-branes probing a $\bC^2/\bZ_k$ singularity compactified on a torus. The six-dimensional theories living on the  branes probing the singularity have in  general some global symmetry, which in our case, for general values of $N$ and $k$, is $\fsu(k)\times \fsu(k)\times \fu(1)$. Upon compactification we might turn on fluxes for abelian subalgebras of the global symmetry supported on the torus (see \cite{Chan:2000qc} for early work on the subject). Without the fluxes the theories have an ${\cal N}=2$ supersymmetry and in fact were argued to have infrared free gauge components \cite{Ohmori:2015pia} (see \cite{delZotto:2015rca} for various ways to reduce on a torus without fluxes). Turning the fluxes on we break the supersymmetry to ${\cal N}=1$. The simplicity of the set-up comes about when we turn on fluxes only for sub-groups of $\fsu(k)\times \fsu(k)$. In such a set-up  the compactifications give rise to theories with known Lagrangians. These turn out to be widely studied toric quiver theories, albeit with additional singlet fields. We thus obtain a novel parametrization of such theories labeling them with the number of M5-branes $N$, the order of the orbifold $k$, and the $2k-2$ discrete numbers defining the fluxes through the torus.

The theories in four dimensions are constructed by studying renormalization group (RG) flows of a quiver theory with $\fsu(N)$ gauge nodes which together with the matter fields triangulate the torus and has $k$ gauge groups winding around one of the cycles of the torus. The number of groups winding around the second cycle is related to the total flux through the torus. Turning on vacuum expectation values for baryonic operators in the set-up one obtains theories which correspond to compactifications on a torus with fluxes. The dictionary between the compactifications and the four-dimensional models was suggested in \cite{Gaiotto:2015usa}. For the dictionary to work one needs to introduce singlet fields coupled through superpotential terms to gauge invariant objects. These superpotential terms are in general irrelevant giving rise to free fields in the IR. Thus, although with non-trivial fluxes all the gauge sectors are UV free, there are generally free chiral fields in the IR.

The dictionary is checked in two main ways. First by showing that the anomalies of the compactification deduced by integrating the anomaly polynomial from six dimensions to four are consistent with the four-dimensional construction. Next, the global symmetry of the theory in four dimensions can be deduced from the compactifiaction details and we give examples of how this works.

In addition to fluxes for continuous symmetries we can turn on fluxes for discrete symmetries of the six-dimensional model. The global structure of the flavor symmetry is $(\SU(k)\times \SU(k)\times \U(1))/{\mathbb Z}_k$. Turning appropriate Steifel-Whitney classes we obtain a larger class of models in four dimensions. These fluxes can materialize in different ways. One way is through fractional fluxes whose quantization is consistent only for $(\SU(k)\times \SU(k))/{\mathbb Z}_k$. Another is by switching on almost commuting holonomies around the cycles of the torus, in the sense that the holonomies commute in $(\SU(k)\times \SU(k))/{\mathbb Z}_k$ but do not in $\SU(k)\times \SU(k)$. In four dimensions this procedure corresponds to constructing the torus by gluing a triangulated cylinder with a twist.

We also discuss the field theories one obtains with fluxes for all possible $\fu(1)$ subgroups in the special case of two M5-branes on ${\mathbb  Z}_2$ singularity where the field theoretic construction is known. Finally we give a prediction for anomalies of theories obtained from six dimensions for general choices of Riemann surfaces. We have no field theoretic constructions in this case and this will serve as a prediction to be contrasted with future computations.

\

\

The paper is organized as follows. In section \ref{sec6d}, we discuss the computation of the anomalies from the six-dimensional vantage point. We consider the anomaly polynomial of $N$ M5 branes probing a $\bC^2/\bZ_k$ singularity compactified on a Riemann surface in the presence of general values of fluxes. We then derive the anomaly polynomial for the four-dimensional models. The case of a torus is discussed in much detail.  In section \ref{sec:4d}, we consider the construction in four dimensions which should result in theories corresponding to torus compactifications. We compute the anomalies and see the agreement with the six-dimensional predictions. In section \ref{sec:symmetry}, we detail several examples deriving precise quiver diagrams and discussing symmetry properties which consistently enhance to match expectations from six dimensions. In section \ref{sec:global}, we discuss compactifications with Steifel-Whitney classes and the four-dimensional theories related to these. We have two appendices: In appendix \ref{sec:trinion}, we deduce some predictions from six dimensions for anomalies of four-dimensional SCFTs. 
The appendix \ref{sec:appendix} details field theoretic constructions of strongly coupled models corresponding to compactifications with general fluxes.

\section{Anomalies from 6d}\label{sec6d}

We begin our discussion from the six-dimensional perspective. Consider taking $N$ M5 branes which probe a $\bC^2/\bZ_k$ singularity and compactify them on a torus. The theory in six dimensions has $\fsu(k)_b\times \fsu(k)_c \times \fu(1)_s$ symmetry for general value of $k$ and $N$; here two $\fsu(k)$ symmetries come from the gauge fields on the $\bC^2/\bZ_k$ singularity on the left and the right of the M5 branes, and $\fu(1)_s$ comes from the isometry of the $\bC^2/\bZ_k$ singularity.\footnote{For $k=2$ the global symmetry is enhanced to $\fsu(2)^3$, as the isometry of $\bC^2/\bZ_2$ is $\fsu(2)$. For $N=2$ the global symmetry is enhanced to $\fsu(2k)$. For $k=N=2$ the global symmetry is enhanced to $\frak{so}(7)$. For $N=1$, we have $k^2$ six dimensional hypermultiplets which transform as a bifundamental representation of the global symmetry. 
}
Upon compactification we can choose an abelian subalgebra of this symmetry and turn on fluxes supported on the torus (see for example \cite{Gaiotto:2015usa,Razamat:2016dpl}). As the first Chern classes of the fluxes have to be properly quantized, the choice gives us models in four dimensions which are labeled by discrete parameters. We can compute the 't Hooft anomalies of the theories from the compactification setup by taking the anomaly eight-form polynomial and integrating this over the torus with the fluxes turned on. This provides a prediction for the four-dimensional models which we will now deduce.

\subsection{Anomaly polynomial of the 6d theory}

  Let us first consider the anomaly coefficients of the six-dimensional $\CN=(1,0)$ theory 
  which is a $\bZ_k$ orbifold of the $A_{N-1}$ type \Nequals{(2,0)} theory.
  These are packaged into the anomaly eight-form polynomial $I_8$, and can be computed using the methods developed in \cite{Ohmori:2014kda,Intriligator:2014eaa}, using the fact that on the tensor branch this theory becomes a linear quiver gauge theory with gauge group $\SU(k)^{N-1}$. 
    We use the normalization where the bifundamental hypermultiplets in the quiver have charge $\pm1$ under $\fu(1)_s$.
  The resulting anomaly polynomial is as follows:
    \bea
    I_8
    &=&   \frac{k^2 N^3}{24} c_2(R)^2 - \frac{N(k^2-1)}{48} c_2(R) (4 c_2(R) + p_1(T))
              \nonumber \\
   & &    - \frac{Nk}{8} c_2(R) (\Tr F_b^2 + \Tr F_c^2)
            + \frac{N}{8} \left( \frac{1}{6}c_2(R)p_1(T) - \frac{1}{6} p_2(T) + \frac{1}{24} p_1(T)^2 \right) 
            \nonumber \\
   &  &   - \frac{1}{2} (I_\text{vec}(b) + I_\text{vec}(c)) - I_\text{tensor} - \frac{1}{2N} \left( \frac{1}{4} \Tr F_b^2 - \frac{1}{4} \Tr F_c^2 \right)^2
           \nonumber \\
   &  &  - N \left(\frac{k}{8} \left( \Tr F_b^2 + \Tr F_c^2 \right) c_1(s)^2
             + \frac{1}{48} k^2 p_1(T)  c_1(s)^2 \right)
   \nonumber \\
   & & + \frac{k}{6} c_1(s) (\Tr _\text{fund} F_b^3 - \Tr _\text{fund} F_c^3)+\frac{k^2  \left(N-N^3\right)}{12} c_2(R) c_1(s)^2+\frac{k^2 N^3}{24} c_1(s)^4,
   \label{6danom}
    \eea
  where $p_1(T)$ and $p_2(T)$ are the first and second Pontryagin classes of the tangent bundle,
  $c_2(R)$ and $c_1(s)$ are the second and the first Chern classes of the $\fsu(2)_R$ and of the $\fu(1)_s$ bundles of the 6d theory, respectively, $\Tr F_b^n$ and $\Tr F_c^n$ are parametrized below by Chern roots, and 
    \begin{align}
    I_\text{vec}(b)&=-\frac1{24}(k^2-1)c_2(R)^2-\frac1{48}(k^2-1)c_2(R)p_1(T) -\frac{k^2-1}{5760}(7p_1(T)^2-4p_2(T)) \nonumber \\
    & \quad 
    -\frac k4 c_2(R) \Tr F_b^2 - \frac k{48} p_1(T) \Tr F^2_b -\frac1{16} (\Tr F_b^2)^2 -\frac k{12} \Tr_\text{fund} F^4_b, \\
    I_\text{tensor} &=\frac1{24}c_2(R)^2 + \frac1{48} c_2(R)p_1(T) + \frac{23}{5760}p_1(T)^2 -\frac{29}{1440} p_2(T).
    \end{align}

\subsection{Mapping the charges in 6d, 5d and 4d}
\label{subsec:charge}

We will match anomalies for various symmetries by performing computations in different dimensions, thus we will start by matching the symmetries between different dimensions.
Let us map here the charges from 6d to lower dimensions.

\paragraph{6d:}
 We have $\fsu(k)_b\times \fsu(k)_c\times \fu(1)_s$ as the flavor symmetry,
in addition to the $\fsu(2)_R$ symmetry. Recall that for $N=1$, we have $k^2$ six dimensional hypermultiplets which transform as a bifundamental representation of the global symmetry.
Let us say that in six dimensions the flavor symmetry bundles split, and the Chern roots are given by 
\begin{equation}
b_1,\ldots, b_k ;~ 
c_1,\ldots, c_k ;~
s
\end{equation} 
  respectively, with $\sum b_j=\sum c_j=0$, so that a component of the bifundamental couples to the line bundle with the Chern class $a_i - b_j  + s$.  
  The Chern class $c_1(s)$ used in \eqref{6danom} is identified with this $s$.
  In our normalization, \bea
    \Tr F_b^2
     =   - 2\sum_i b_i^2, ~~~
    \Tr _\text{fund} F_b^4
     =     \sum_i b_i^4, ~~~
    {\rm \Tr}_\text{fund} F_b^3
     =     \sum_i b_i^3,
    \eea
  and 
    \begin{equation}
    c_2(R)=-x^2,
    \end{equation} 
for the $\fsu(2)_R$ bundle with Chern roots $(x,-x)$. 

\paragraph{5d:}
Let us put the 6d theory on ${\mathbb S}^1$ with a nontrivial holonomy for the flavor symmetry.
Then in the infrared, the 5d theory is dual, in the sense described in \cite{Aharony:1997bh} as continuation past infinite coupling, to the circular quiver $\fsu(N)^k$.
Call $I_i$ the Chern root for the instanton number symmetry of $\fsu(N)_i$ and
$t+H_i$ the Chern root for the baryon number symmetry of the bifundamental between $\fsu(N)_i$ and $\fsu(N)_{i+1}$, with $\sum H_i=0$. 
From the discussion in \cite{Tachikawa:2015mha}, 
we know that the one-instanton operator of $\fsu(N)_i$, that becomes the raising/lowering operators of the $\fsu(k)^2$ flavor symmetry currents, couples to a line bundle with the Chern class \begin{equation}
I_i \pm \frac12 (N H_i - N H_{i-1}).
\end{equation} This is to be identified with $b_i-b_{j-1}$ and $c_i-c_{j-1}$. Therefore, we see\begin{equation}
N H_i= b_i-c_i.\label{q}
\end{equation}

Next, to relate $t$ and $s$, it is useful to consider the Higgs branch of the theory, when we separate $N$ M5-branes. In the following, we will write down some key invariants on the Higgs branch and specify some of the relations they satisfy. This will be sufficient to derive the quantum numbers.
In the 5d description, one can use the Kronheimer construction for the $N$-th symmetric product of $\bC^2/\bZ_k$: denoting the bifundamental between $\fsu(N)_i$ and $\fsu(N)_{i+1}$ as $\Phi_i$ and $\tilde \Phi_i$, 
let $\diag(z_a)=\Phi_i \tilde \Phi_i$, $\diag(x_a)=\Phi_1\cdots \Phi_k$, $\diag(y_a)= \tilde\Phi_1\cdots\tilde\Phi_k$. They satisfy $x_ay_a = z_a^k$ (with no sum on $a$). 
In terms of $(u,v)\in \bC^2$, we have $x=u^k$, $y=v^k$, $z=uv$.
Now, $x$ and $y$ have the $\fu(1)_t$-charge $\pm k$. Therefore, $u$ and $v$ have the $\fu(1)_t$-charge $\pm 1$. 

In six dimensions, again when $N$ M5-branes are separated, the same Higgs branch can be found as explained in \cite{Ohmori:2015pua}.
Namely, when we denote the $\fsu(k)^2$ bifundamental  by $\Phi$, $\tilde \Phi$, we have $z=\Tr \Phi\tilde \Phi$, $x=\det \Phi$, and $y=\det \tilde \Phi$ so that $xy=z^k$. We had declared that $\Phi$, $\tilde \Phi$ have $\fu(1)_s$ charge $\pm1$,
thus $x,y$ have charge $\pm k$, thus $u,v$ have charge $\pm1$. 
Therefore, we can equate the $\fu(1)_s$ charge and $\fu(1)_t$ charge: \begin{equation}
s=t.
\end{equation}

The $\fsu(2)_R$ symmetry in 5d and 6d can be naturally identified so the scalars in the bifundamental hypermultiplets in the $\fsu(N)^k$ quiver are $\fsu(2)_R$ doublets. 

\paragraph{4d:}
Now we consider the situation in four dimensions. 
In the tube theory, most of the analysis above can be directly applied. The $\fsu(2)_R$ symmetry is broken to the Cartan. We use the normalization where the supercharge has the charge $\pm1$ under the remaining $\fu(1)_{R'}$ symmetry. Here $R'$ emphasizes that this is a natural $R$-symmetry coming from the six-dimensional construction; this generically will \emph{not} be the superconformal $R$ symmetry in the infrared, which needs to be determined by the $a$-maximization \cite{Intriligator:2003jj}.

In any case, the bifundamentals in the $\fsu(N)^k$ quiver, before the supersymmetry is broken by half, have the $\fu(1)_{R'}$-charge $1$
and the $\fu(1)_t$-charge $\pm1$. 
Then, the surviving chiral bifundamental in the $\fsu(N)^k$ tube theory 
has the $\fu(1)_{R'}$-charge $1$ and the $\fu(1)_t$-charge $1$.
Together with \eqref{q}, this data on the tube theory is enough to find the charge assignment in the Lagrangian class ${\cal S}_k$ theory, as we will see in the next section.

\subsection{Anomaly polynomial from 6d}
  We now compute the anomaly polynomial of the compactified theory from the 6d point of view.
  Let $N_{b_{i}}$, $N_{c_{i}}$, and $N_s$ be the numbers of fluxes 
  of the $\fu(1)_{b_{i}}$, $\fu(1)_{c_{i}}$, and $\fu(1)_s$ respectively.
  Let us also denote the first Chern classes of line bundles in 4d as $c_{1}(R')$, $c_1 (t)$, $c_1(\beta_i)$ and $c_1(\gamma_i)$.
  The Chern roots introduced above are related as follows\footnote{When $g=1$, $\frac{t}{2g-2}$ should mean a two-form whose integral over the Riemann surface is $-1$. The proceeding equations hold also for this case.}
    \begin{align}
    x
    &=  c_{1}(R') - \frac{t}{2}, &
    s
     &=    c_1 (t)  + N_s \frac{t}{2g-2} ,
           \nonumber \\
    b_{i}
    &=    N c_{1}(\beta_{i}) - N_{b_{i}} \frac{t}{2g-2},  &
    c_{i}
     &=     N c_{1}(\gamma_{i}) - N_{c_{i}} \frac{t}{2g-2},
           \label{symmetry}
    \end{align}
  where $c_2(R) = - x^2$, $c_2(s) = - c_1(s)^2 = - s^2 $ and $\int_{C_g} t = 2-2g$.
  
  By substituting these into the anomaly eight-form and performing the integral over the Riemann surface $\int_{C_g} I_{8}$, we get
    \bea\label{eighf}
    I_{6}
   &=&  \frac{(k^2-2)(N-1)}{48} (2g-2) c_1(R') p_1(T_4) + \frac{k^2 N}{24}  N_s c_1 (t) p_1(T_4)
             \nonumber \\
    &+&  \frac{(N-1)(k^2(N^2+N-1)+2)}{12} (2g-2) c_1(R')^3
            + \frac{k^2 N(N^2-1)}{6} N_s c_1(R')^2 c_1(t)
               \nonumber \\
    &-&  \frac{k^2 N(N^2-1)}{12}(2g-2) c_1(R') c_1(t)^2
            - \frac{k^2 N^3}{6} N_s c_1(t)^3 
              \nonumber \\
    &+&   \frac{1}{2} \left( -k N (N-1) c_1(R')^2 + kN^2 c_1 (t)^2 \right) \sum_i (N_{b_i} c_1(\beta_i)  + N_{c_i} c_1(\gamma_i))
            \nonumber \\
    &-&  \frac{kN^2(N-1)}{4} (2g-2) c_1(R') \sum_{i} \left( c_{1}(\beta_{i})^2 + c_{1}(\gamma_{i})^2 \right) - \frac{kN}{24} \sum_{i} \left( N_{b_{i}} c_{1}(\beta_{i}) + N_{c_{i}} c_{1}(\gamma_{i}) \right) p_{1}(T_{4})
    \nonumber \\
    &-&   \frac{kN^2}{2} \sum_i \left(  (N N_s - N_{b_i}) c_1 (t) c_{1} (\beta_{i})^2 +  (N N_s + N_{c_i}) c_{1}(t) c_{1}(\gamma_{i})^2 \right)           \nonumber \\
    &+&   \frac{N^{2}(N-1)}{2} (\sum_{i}c_{1}(\beta_{i})^{2})(\sum_{j} N_{b_{j}} c_{1}(\beta_{j}))
         + \frac{N^{2}(N-1)}{2} (\sum_{i}c_{1}(\gamma_{i})^{2})(\sum_{j} N_{c_{j}} c_{1}(\gamma_{j}))
           \nonumber \\
    &+&   \frac{kN^{3}}{6} \sum_{i}((N_{b_{i}}-N_s) c_{1}(\beta_{i})^{3} + (N_{c_{i}}+N_s) c_{1}(\gamma_{i})^{3})
     \nonumber \\
    &+&   \frac{N^{2}}{2} \Bigg( (\sum_{i}c_{1}(\beta_{i})^{2})(\sum_{j} N_{c_{j}} c_{1}(\gamma_{j}))
         + (\sum_{i}c_{1}(\gamma_{i})^{2})(\sum_{j} N_{b_{j}} c_{1}(\beta_{j})) \Bigg),
    \eea
  where $\sum_{i=1}^{k}c_1(\beta_i)=0$ and $\sum_{i=1}^{k}c_1(\gamma_i)=0$.
  This is the anomaly six-form of the compactified 4d theory.

  Let us now give an expression for a torus without $\fu(1)_s$ flux.
  By setting $g=1$ we get the following anomaly polynomial:
    \bea
    I_{6}
    &=&   \frac{1}{2} \left( -k N (N-1) c_1(R')^2 + kN^2 c_1 (t)^2 \right) \sum_i (N_{b_i} c_1(\beta_i)  + N_{c_i} c_1(\gamma_i))
            \nonumber \\
    & &  - \frac{kN^2}{2} \sum_i \left(  -N_{b_i} c_1 (t) c_{1} (\beta_{i})^2 + N_{c_i} c_{1}(t) c_{1}(\gamma_{i})^2 \right) 
           - \frac{kN}{24} \sum_{i} \left( N_{b_{i}} c_{1}(\beta_{i}) + N_{c_{i}} c_{1}(\gamma_{i}) \right) p_{1}(T_{4})
             \nonumber \\
    & &  + \frac{N^{2}(N-1)}{2} (\sum_{i}c_{1}(\beta_{i})^{2})(\sum_{j} N_{b_{j}} c_{1}(\beta_{j}))
         + \frac{N^{2}(N-1)}{2} (\sum_{i}c_{1}(\gamma_{i})^{2})(\sum_{j} N_{c_{j}} c_{1}(\gamma_{j}))
           \nonumber \\
    & &  + \frac{kN^{3}}{6} \sum_{i}(N_{b_{i}} c_{1}(\beta_{i})^{3} + N_{c_{i}} c_{1}(\gamma_{i})^{3})
     \nonumber \\
    & &  + \frac{N^{2}}{2} \Bigg( (\sum_{i}c_{1}(\beta_{i})^{2})(\sum_{j} N_{c_{j}} c_{1}(\gamma_{j}))
         + (\sum_{i}c_{1}(\gamma_{i})^{2})(\sum_{j} N_{b_{j}} c_{1}(\beta_{j})) \Bigg).
    \eea
  where $\sum_{i=1}^{k}c_1(\beta_i)=0$ and $\sum_{i=0}^{k-1}c_1(\gamma_i)=0$.
  
  The triangle anomaly in the more traditional form $\Tr xyz$ can be read off by first substituting  $c_1(\beta_k) = -\sum_{i=1}^{k-1}c_1(\beta_i)$ and the corresponding expression for $c_1(\gamma_k)$, and then reading off the coefficient of $c_1(x)c_1(y)c_1(z)$. Just to give an example, we have $\Tr t^2 \beta_i= kN^2 (N_{b_i} - N_{b_k})$.

\section{4d theories from tori}
\label{sec:4d}

\subsection{Structure of the 4d quiver theories}
\begin{figure}[t]
\center
\centerline{\includegraphics[scale=0.7]{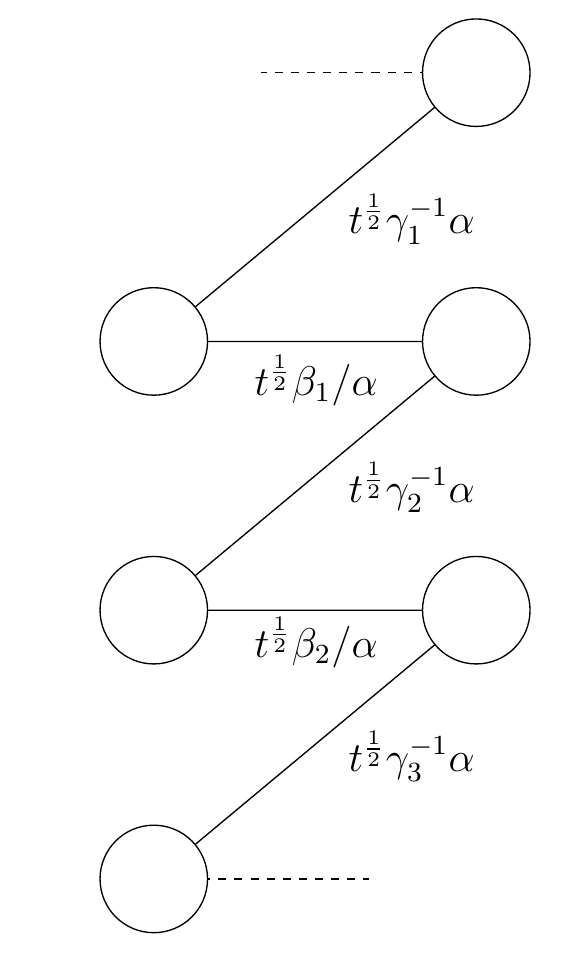}\;\;\;\;\;\;\;\;\;\,\,\,\,\,\,\, \includegraphics[scale=0.6]{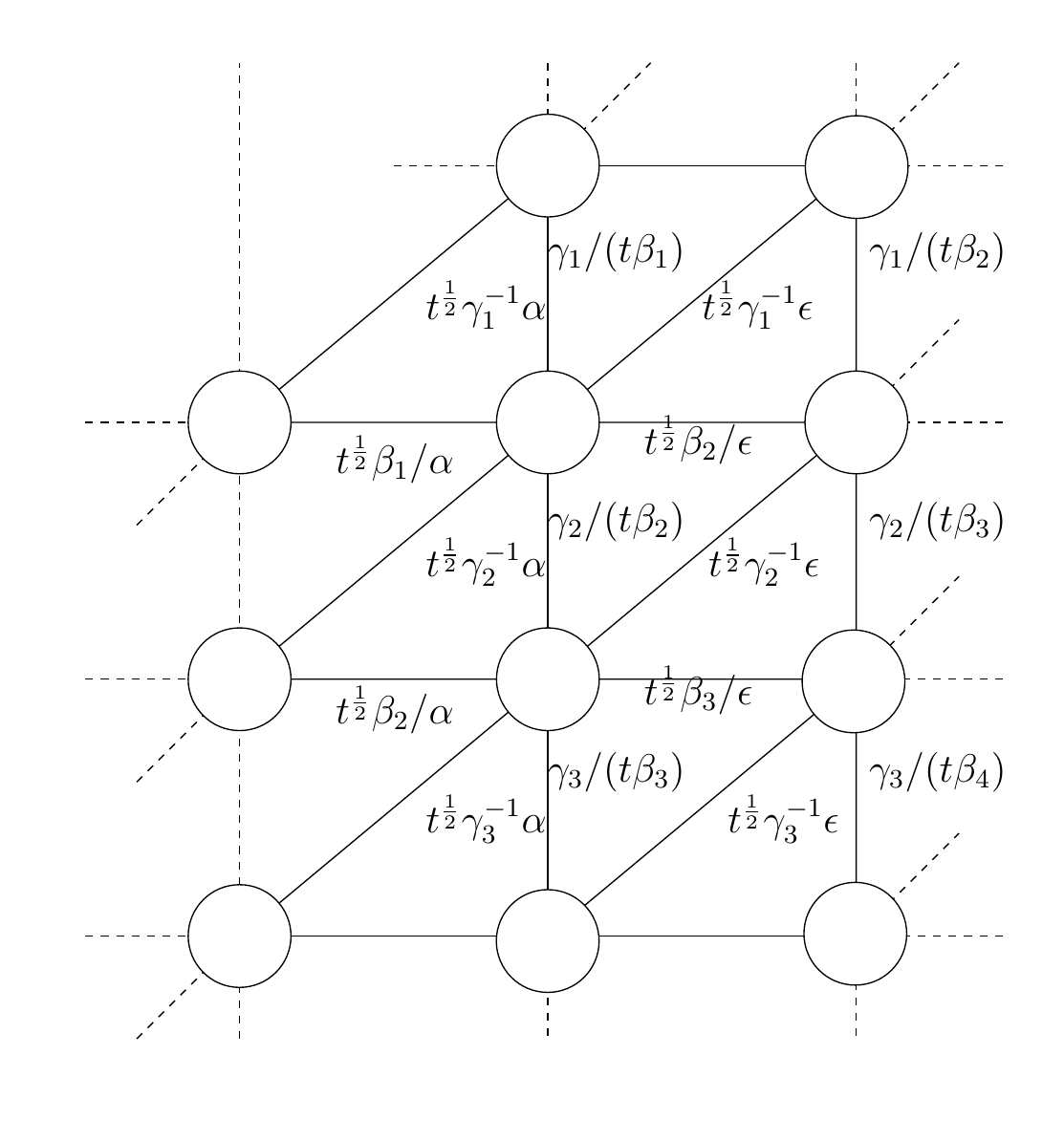} }
\caption{On the left we have the free trinion. This is a collection of $N^2(2k)$ free fields. We organize these as bifundamentals of two copies of $\fsu(N)^k$. In the picture the circles are $\fsu(N)$ groups and one has $k$ groups winding around a cylinder. The trinion is associated to a compactification on a sphere with two maximal punctures (of different color) and a minimal puncture. On the right we glue trinions together to triangulate a torus. We have $lk$ trinions combined with every $\Phi$ gluing introducing bi-fundamental fields which appear as vertical lines in the diagram. }
\label{qtr}
\end{figure}

We consider now the field theory construction corresponding to tori with fluxes with no punctures~\cite{Gaiotto:2015usa}.  The models are constructed by first starting from a toric quiver built from some number of free trinion theories (figure \ref{qtr}),
and then by higgsing some of the symmetries.
 
The free trinion theory corresponds to a sphere with three punctures: two are maximal and one is minimal, and the $\fu(1)_{s}$ flux $N_s = \half$ (in the normalization used in ~\cite{Gaiotto:2015usa,Razamat:2016dpl}).
The maximal and the minimal punctures are associated with $\fsu(N)^k$ and $\fu(1)$ flavor symmetries respectively.
The former is known to be labeled by the color $c \in \mathbb{Z}_k$ and the sign $\sigma$ $(=+1, -1)$ \cite{Gaiotto:2015usa,Hanany:2015pfa}; 
The two maximal punctures associated to the free trinion in figure \ref{qtr} have the same signs $\sigma = +1$.
We assign the canonical $R'$-charge $\half$ to all the fields, and the other charges are denoted in figure \ref{qtr}.

Gluing of two maximal punctures corresponds to gauging of the $\fsu(N)^k$ symmetry of both punctures. Depending on the signs of the punctures we have two gluings \cite{Gaiotto:2015usa,Franco:2015jna,Hanany:2015pfa}:
  \begin{itemize}
  \item $\Phi$ gluing: when the two punctures have the same sign, say $\sigma = +1$, we add an $\CN=1$ $\fsu(N)^k$ vector multiplet and 
  bifundamental chiral multiplets between these gauge factors cyclically, with superpotential coupling of the bifundamentals and the mesonic operators associated to the punctures; the quiver in figure \ref{qtr} represents $\Phi$ gluing.
  \item $S$ gluing: when the two punctures are different, we only add an $\CN=1$ $\fsu(N)^k$ vector multiplet
  with the superpotential coupling of two mesonic operators coming from the two punctures.
  \end{itemize}
These are associated to the theories on a tube without any flux. (See Appendix for the tube theory with $\fu(1)_s$ flux.) 
As already noticed in \cite{Razamat:2016dpl} one can see that the charge assignment is consistent with the discussion in section \ref{subsec:charge} from 6d.
If the two punctures have the same color this preserves all internal symmetries. However if the two punctures have different colors then all $\fu(1)_{\beta}$'s are broken.

We here only focus on the $\Phi$ gluings and construct a quiver theory
associated to a torus with only minimal punctures from a collection of $k l$ free trinions, as in figure \ref{qtr}.  
When $l$ is an integer we can always glue two punctures that have the same color preserving all the internal symmetries, so the global symmetry of this model consists of $\fu(1)_\beta^{k-1}\times \fu(1)_\gamma^{k-1}\times \fu(1)_t$ and $kl$  $\fu(1)_{\alpha_j}$ symmetries which are associated to minimal punctures.

\begin{figure}[t]
\center
\includegraphics[scale=0.53]{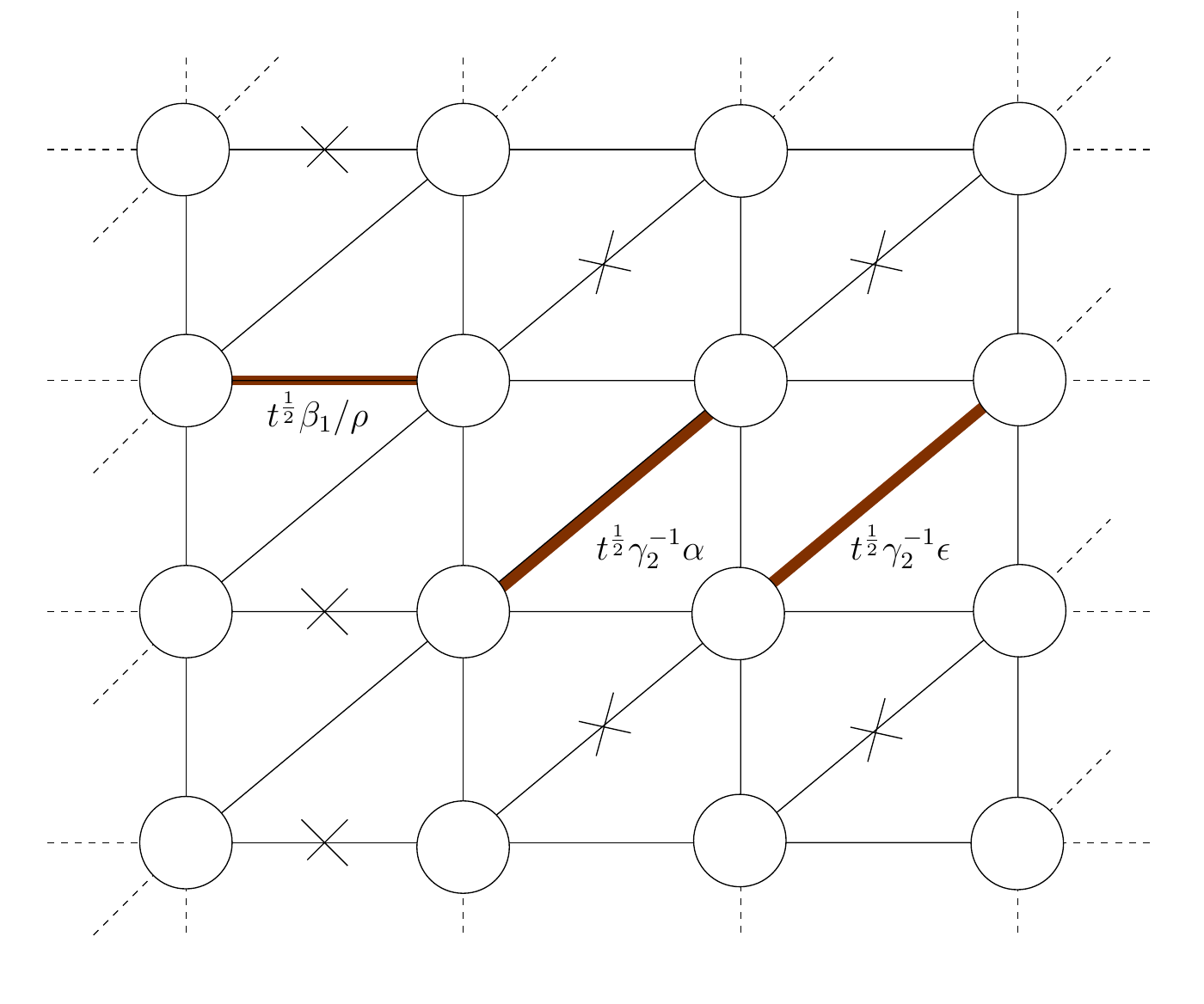}
\caption{We get rid of all the minimal punctures by giving vacuum expectation values to baryons. In every free trinion we give a vacuum expectation value to one of the baryons. The choice of the baryons is related to the flux and in general different choices lead to different theories in the IR. The baryons which do not receive a vev but are charged  with same charge under the minimal puncture symmetry as the baryon which does receive vev, are flipped. In the diagrams the fields with a cross are the baryons which are flipped. In the picture the baryons which receive vacuum expectation value are weighed as $t^{\frac{N}2}\beta_1^N/\rho^N$, $t^{\frac{N}2}\gamma_2^N\alpha^N$, $t^{\frac{N}2}\gamma_2^N\epsilon^N$.
} \label{fig:quive}
\end{figure}

We obtain models with no punctures by giving vacuum expectation values to $kl$ baryonic operators charged under $\fu(1)_{\alpha_j}$ symmetries and introducing certain gauge-singlet chiral multiplets flipping some of the other baryons\footnote{By flipping an operator ${\cal O}$ we mean the procedure of adding a chiral field $M_{\cal O}$ to the model with superpotential $W=M_{\cal O}\;\, {\cal O}$.}, as sketched in figure \ref{fig:quive}. There are choices to be made as to which baryons the vacuum expectation values are given and this choice maps to a choice of fluxes in six dimensions \cite{Gaiotto:2015usa}. We will write down the exact correspondence shortly. 
This procedure produces $\fu(1)_s$ flux $N_s = - \half$. 
After higging all the $\fu(1)_{\alpha_j}$ symmetries, the $\fu(1)_{s}$ flux produced is cancelled by the $N_s =k l$ flux of the previous theory.
Then the resulting theory corresponds to a torus with $N_s = 0$. 
With $\fu(1)_s$ flux the models do not in general have known regular Lagrangians (see for example the discussion in the case of $k=1$ in \cite{Bah:2012dg,Fazzi:2016eec,Nardoni:2016ffl}). We will discuss one case with $\fu(1)_s$ flux in   appendix \ref{sec:appendix}.

Let us make  several general observations about these models. 
The quivers  correspond to tiling of the torus with triangular and square faces. The exact details depend on the 
fluxes and in four-dimensional language on the ways we close the minimal punctures. The theory with the minimal punctures we utilize as a starting point of the construction triangulates the torus. Importantly we supplement the quiver with singlet fields, some of which might be free and some coupled to gauge singlet combinations of fields through additional superpotential terms.

Such theories were widely
studied in various contexts about ten years ago~\cite{Hanany:2005ve,Franco:2005rj,Franco:2005sm}.
It is convenient to think about the theories in terms of zigzag paths on the torus. Each symmetry factor $\fu(1)$ (with exception of $\fu(1)_t$) corresponds to a loop, zigzag path, winding around the cycles of the torus. Let us call the cycle around which we have, in the theory in the UV, $k$ gauge groups cycle A and the other one cycle B.  The UV model then has loops winding once around cycle A corresponding to puncture symmetries and $k$ loops winding $l$ times around each one of the cycles A and B, and $l$ loops winding $-l$ times around cycle A and $l$ times around cycle B. The  flow initiated by closing the punctures preserves the symmetries not associated to the punctures and breaks symmetries associated to punctures. The pattern of winding of the different lines can be translated to the fluxes. For example, the torus with no flux is formally mapped to configurations with all windings vanishing. 

\subsection{Anomalies of the 4d quiver theories}

The anomalies for these models can be rather easily derived.
For the sake of computation of the anomalies we do not need to figure out the quiver diagram in the IR of the flow triggered by vacuum expectation values turned on for baryonic operators when closing minimal punctures. We can compute these in the UV, making sure to use the symmetries surviving in the IR and decoupling the relevant Goldstone chiral multiplets.
Let us give the algorithm for computing the anomalies. We will encode all anomalies in the  trial $a$ conformal central charge and in the trace of a trial $R$ symmetry where we will keep  dependence on possible mixing parameters with all the abelian symmetries. We denote the $R$ charge as
\be
R(s^\beta,s^\gamma,s_t,s^\alpha)=R'+\sum_{i=1}^k s^\beta_i q_i^\beta+\sum_{i=1}^k s^\gamma_i q_i^\gamma+s_t q_t+\sum_{u=1}^{kl} s^\alpha_u q^\alpha_u\,,
\ee 
  where $q_i^\beta$, $q_i^\gamma$, $q_t$ and $q_u^\alpha$ are the charges of $\fu(1)_\beta^{k-1}$, $\fu(1)_\gamma^{k-1}$, $\fu(1)_t$
  and $\fu(1)_{\alpha_j}$.
  We have the constraint $\sum_{j=1}^ks^\beta_j=\sum_{u=1}^k s^\gamma_u=0$. 
  The conformal anomaly $a$ of a free chiral multiplet of $R$-charge $z$ and of a vector multiplet of group of dimension $h$ are given by
\be a_\chi(z)=\frac3{32}(3(z-1)^3-(z-1))\,,\;\;\;\;\;\;\;\; a_v(h)= \frac3{16}h\,.\ee 
Then the anomaly of the free trinion is
\be
a_t(N,k;s^\beta,s^\gamma,s_t,s_\alpha) 
&=&N^2\sum_{j=1}^ka_\chi(\frac12+s^\beta_j+\frac12 s_t+s_\alpha)+N^2\sum_{j=1}^ka_\chi(\frac12-s^\gamma_j+\frac12 s_t-s_\alpha)
\nonumber\\
&=& \frac3{32}N^2(k(1- s_t)+\frac34 k (s_t-1)^3+9 k(s_t-1)s_\alpha^2\\
&&+3(-S^\gamma_3+S^\beta_3+
\frac32(s_t-1)(S^\gamma_2+S^\beta_2)+3s_\alpha(S^\beta_2-S^\gamma_2)))\,.\nonumber
\ee 
where
\be
S^\beta_i=\sum_{u=1}^k (s^\beta_u)^i\,,\qquad\; S^\gamma_i=\sum_{u=1}^k (s^\gamma_u)^i\,.
\ee 
 We note that the color of the maximal punctures is not important in the computation of the anomaly of the free trinion as it only determines the sequence in which the different chiral fields are organized together.
 
The anomalies of each $\Phi$ gluing of maximal punctures of color $c$ are easily seen to be
\be
&&a_g(N,k,c;s^\beta,s^\gamma, s_t)=k a_v(N^2-1) + N^2\sum_{i=1}^k a_\chi(1-s_t-s^\beta_i+s^\gamma_{i+c})\\
&&~~~~~~~~~~~ =\frac3{16} k(N^2-1)+\frac3{32}N^2(k s_t-3ks_t^3-3\sum_{u=1}^k((s^\beta_u-s^\gamma_{u+c})^3+3s_t
(s^\beta_u-s^\gamma_{c+u})^2))\,,
\nonumber
\ee where the indices are summed mod $k$. These fields are charged under both $\beta$ and $\gamma$ symmetries and thus the color of the puncture we glue is important for the anomaly.
The anomaly of torus built from $k l$ free trinions is then
\be
a_{tor}(N,k,l,s^\beta,s^\gamma,s_t,\{s_\alpha\}) 
&=& \sum_{i=1}^{l k} a_t(N,k; s^\beta,s^\gamma,s_t,s_{\alpha_i})+l(\sum_{i=1}^ka_g(N,k,i;s^\beta,s^\gamma,s_t)) \nonumber \\
&=&\frac{3}{128} k^2 l \left(9 N^2 (1-s_t^2) (s_t+1)-8\right)-\frac{27}{64}N^2 k(s_t+1) l(S^\gamma_2+S^\beta_2)\nonumber\\
&+&\frac{27}{32}(s_t-1)N^2k\sum_{j=1}^{k\;\,l}(s_{\alpha_j})^2+\frac{27}{32}N^2(S^\beta_2-S^\gamma_2)  \; \sum_{j=1}^{k\;\,  l} s_{\alpha_j}\,.
\ee 

To close minimal punctures we need to substitute $s_{\alpha_i}=-\half-\frac12s_t-s^\beta_{F(i)}\equiv {\frak s}_i$ or $s_{\alpha_i}=\half+\frac12s_t-s^\gamma_{F(i)}\equiv {\frak s}_i$ and flip the baryons. The specification of the fugacities reflects the fact that the operators receiving the vacuum expectation values have all their charges vanishing in the IR. The function $F(i)$ is an arbitrary function mapping $(1,\cdots,k)$ to itself.
  We denote by $Q^\beta_i$ the number of minimal punctures closed with $s^\beta_i$, and $Q^\gamma_i$ is the same regarding $\fu(1)_{\gamma_i}$. We denote by ${\frak Q}^\beta=\sum_{i=1}^k Q^\beta_i$, ${\frak Q}^\gamma=\sum_{u=1}^k Q^\gamma_u$. 

The conformal anomaly $a$ for the torus with no punctures but with fluxes determined by a choice of $F$ is then
\be
&&a_{tor, F}(N,k,Q^\beta_i,Q_j^\gamma,s^\beta,s^\gamma,s_t)=a_{tor}(N,k,\frac{{\frak Q}^\beta+{\frak Q}^\gamma}k,s^\beta,s^\gamma,s_t,s_{\alpha_i}\to {\frak s}_i) \nonumber\\
&&~~~~~~~~~~~~~~~+\sum_{i=1}^k\sum_{u=1}^k Q^\gamma_u a_\chi(2-N(s^{\gamma}_u-s^{\gamma}_i))+\sum_{i=1}^k\sum_{u=1}^k Q^\beta_u a_\chi(2-N(s^{\beta}_{i}-s^{\beta}_u))
\ee where in the second line we have the contribution of the chiral fields which flip the baryons and in the first line we specialize the parameters of the torus with punctures to be consistent with vacuum expectation values.  Note that in the second line the terms with $u=i$ subtract the Goldstone bosons appearing in the flow as we break some  symmetries.
This evaluates to 
\begin{equation}
\begin{aligned}
=& (\frac{27}{32}N^2k(1-s_t^2)-\frac{27}{32}N^2(S^\beta_2-S^\gamma_2)-\frac34 kN -\frac{27}{32}N^3 S^\gamma_2)\langle Q^\gamma, s^\gamma\rangle \\
& - (\frac{27}{32}N^2k(1-s_t^2)-\frac{27}{32}N^2(S^\gamma_2-S^\beta_2)-\frac34 kN -\frac{27}{32}N^3 S^\beta_2)\langle Q^\beta, s^\beta\rangle\\
& +\frac{27}{32}N^2ks_t(\langle Q^\gamma-\frac{{\frak Q}^\gamma}k, (s^\gamma)^2\rangle+\langle Q^\beta-\frac{{\frak Q}^\beta}k,\; (s^\beta)^2\rangle)\\
 &-\frac9{32}kN^3(\langle Q^\gamma-\frac{{\frak Q}^\gamma}k, (s^\gamma)^3\rangle-\langle Q^\beta-\frac{{\frak Q}^\beta}k,\, (s^\beta)^3\rangle)
\,
\end{aligned}\label{trian}
\end{equation}
where $\langle\cdots\rangle$ denotes the inner product of $k$-dimensional vectors and $(s^{\beta,\gamma})^n$ stands for the vector whose components are $(s^{\beta,\gamma}_i{})^n$. 

This expression has several nice features. If one shifts all $Q_i^\beta$ (or $Q^\gamma_i$) by some integer the above does not change.  This corresponds to completely closing minimal punctures with no remaining flux \cite{Gaiotto:2015usa}. Taking $N=2$ we get an expression which is symmetric under exchanging all of $Q^\beta_i-\frac{{\frak Q}^\beta}k$ and $\frac{{\frak Q}^\gamma}k-Q_j^\gamma$. This is consistent with the expectation that the group here is enhanced to $\fsu(2k)$ as below these combinations are identified with the fluxes.

The anomalies here are in agreement with the anomalies computed from six dimensions. The map between the parameters is
\be\label{maraet}
N_{b_i} =Q^\beta_i-\frac{{\frak Q}_\beta}k\,,\qquad
N_{c_i} =\frac{{\frak Q}^\gamma}k-Q^\gamma_i\,,
\qquad N_s=0\,.
\ee

 We can compute the anomaly $b=\Tr R'$ in addition to the trial $a$ anomaly above. For the free trinion we obtain that $b_{trin.}=kN^2(s_t-1)$.  For $\Phi$ gauging we obtain $b_g=k(N^2-1)-N^2k s_t$.
This is independent of the puncture symmetries and the $\beta$ and $\gamma$ symmetries. The anomaly of the torus with $k l$ free trinions is $b_{torus}=-k^2 l$. When we close the minimal punctures we have to flip the baryons and the anomaly is
\be
b(Q^\beta_i,Q^\gamma_i)&=&-k({\frak Q}^\gamma+{\frak Q}^\beta)+k ({\frak Q}^\gamma+{\frak Q}^\beta)-N k \langle Q^\gamma,\; s^\gamma\rangle+Nk\langle Q^\beta\,,\; s^\beta\rangle\\
&=&N k (-\langle Q^\gamma,\; s^\gamma\rangle+\langle Q^\beta\,,\; s^\beta\rangle)\,.\nonumber
\ee
 We observe that with our identification of fluxes with multiplicities of the various choices of closures of minimal punctures all anomalies agree between 4d and 6d.

We can use the trial $a$-anomaly we have obtained to compute the conformal anomalies of the theories. Here we have to be careful as in general the fields which flip the baryons are coupled through irrelevant interactions and thus are free. One then needs to take this into account in the computation of the conformal anomalies.

\

\section{Case studies}
\label{sec:symmetry}

We construct examples of various quiver gauge theories of class ${\cal S}_k$ type associated to a torus for small values of $k$. We will study in greater detail some of their properties. Specifically we calculate the superconformal index and test the global symmetry of the 4d fixed point with that predicted from the 6d construction.

The superconformal index also allows us to compute the dimension of the conformal manifold of the 4d theory. This can also be predicted based on the 6d construction as done for the case of $N=k=2$ case and Riemann surfaces of $g>1$ in \cite{Razamat:2016dpl}. Basically, we expect an exactly marginal operator for each complex structure modulus, and flat connections for the global symmetries. For the case of a torus, we always have a single complex structure modulus. In addition we can also have flat connections for the global symmetries with non-trivial values around each of the two cycles of the torus. These must be abelian due to the homotopy group relation of the torus. Thus we see that we get $2(2k-1)$ real parameters or $2k-1$ complex marginal deformations. So to conclude, we expect:
\be
\dim(\mathcal{M}) = 2k \label{ConfM}.
\ee
We can use the 4d superconformal index to check this prediction.

It is instructive to consider the $k=1$ case, or more specifically, $\mathcal{N}=4$ super Yang-Mills when considered as the $A_{N-1}$ $(2,0)$ theory compactified on a torus. In that case, (\ref{ConfM}) gives a two dimensional conformal manifold which is indeed the known result in $\mathcal{N}=1$ class ${\cal S}$ \cite{Bah:2012dg}. However, it is known that $\fsu(N)$ $\mathcal{N}=4$ super Yang-Mills has in fact $3$ marginal deformations for $N>2$ and just $1$ for $N=2$. Thus we see that this reasoning might be too naive for the case of a torus. 
We start with discussing the $k=N=2$ case, and then move on to discuss the $k=2, N=3$ and $k=3, N=2$ cases.

\

\subsection{$N=k=2$}

Consider taking two free trinions and connecting maximal punctures of the same color together. This results in a torus with two minimal punctures. Then by closing the minimal punctures, we can get a theory corresponding only to a torus, as discussed in the previous section.

\begin{figure}[t]
\center
\centerline{\includegraphics[width=.43\textwidth]{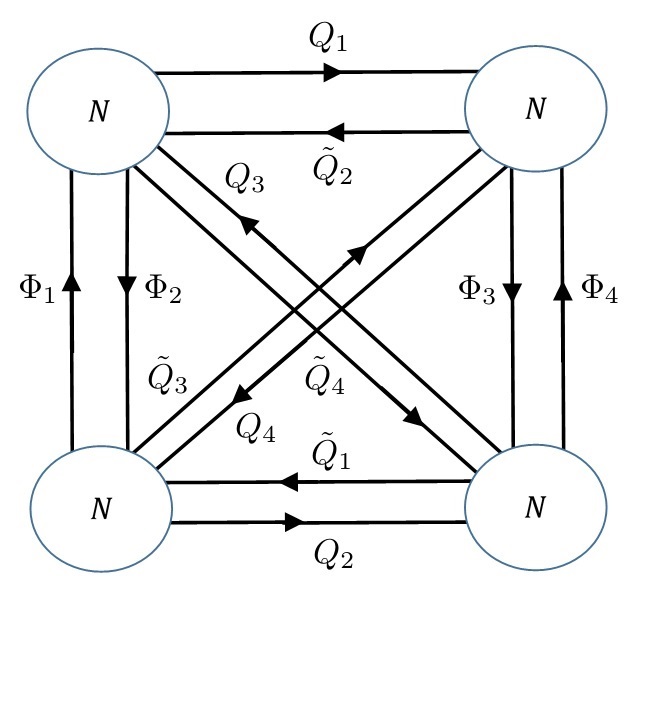}\;\;\;\;\;\;\;\;\;\,\,\,\,\,\,\, \includegraphics[scale=1.1]{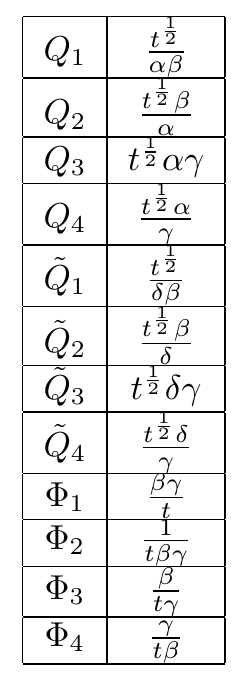} }
\caption{On the left is the quiver diagram for the theory one gets when gluing together two free trions along all maximal punctures for $k=2$. On the right is a table summarizing the charges of the fields under all the non $R$-symmetries: the internal $\fu(1)_{\beta}, \fu(1)_{\gamma}, \fu(1)_{t}$, and the minimal puncture ones $\fu(1)_{\alpha}, \fu(1)_{\delta}$. Additionally there is a cubic superpotential for every triangle. Alternatively it is given by the most general cubic superpotential that is gauge invariant and consistent with the symmetry allocation in the table. All fields have the free $R$-charge $\frac{2}{3}$.}
\label{Stquiver}
\end{figure}

First we begin with the theory corresponding to a torus with two minimal punctures we get by connecting two free trinions. For the purpose of constructing these theories we will leave $N$ general,  setting it to the desired value at the end. The quiver diagram of the theory is shown in figure \ref{Stquiver}. It has a cubic superpotential for any triangle. This is the theory that lives on $N$ D3-branes probing a $\bC^3/\bZ_2 \times\bZ_2$ singularity and has been studied in various contexts. See e.g.~\cite{Feng:2000mi} and references therein.

Let us denote the fluxes of the theory associated with the surface by $(N_b, N_c, N_s)$.
As the free trinion has flux $(0,0,\frac{1}{2})$~\cite{Gaiotto:2015usa,Razamat:2016dpl}, this theory should correspond to flux $(0,0,1)$.

Next we give a vev to the baryon made from $Q_1$. This corresponds to closing a minimal puncture. The resulting theory is associated to a torus with a minimal puncture and flux $(\frac{1}{2},0,\frac{1}{2})$.

Further we can give a vev to another baryon, associated with the other puncture. This will close the other puncture and leads us to a torus with no punctures. We have three distinct choices for the baryon. These will differ by the flux on the torus.

\begin{figure}[t]
\center
\includegraphics[width=0.27\textwidth]{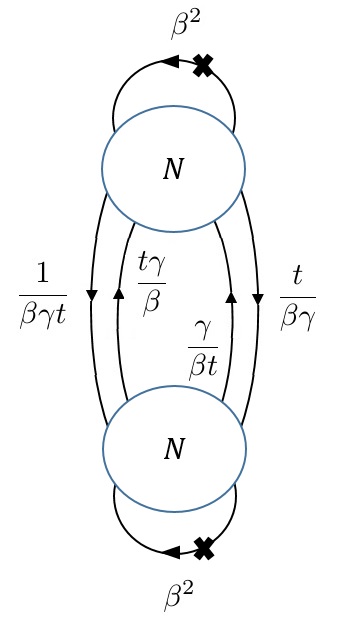} 
\caption{The quiver diagram for the 4d class $\CS_2$ theory corresponding to a torus with flux $(1,0,0)$. Next to the fields are their charges summarized through fugacities. We use mostly standard notation except for two points: lines from a group to itself represent $N^2$ hypermultiplets forming the adjoint plus singlet representations of the group; we write an X over a field to represent the fact that the baryon of that field is flipped. The theory has a cubic superpotential for every triangle which can also be derived by considering the most general cubic superpotential that is gauge invariant and consistent with the symmetry allocation. There is also the superpotential term which is not generally cubic coming from the flipping. All fields, save the flipping fields, have the free $R$-charge $\frac{2}{3}$.}
\label{Affinequiver}
\end{figure}

One choice is to close with the baryon made from $\tilde{Q}_2$. This will lead to a torus with flux $(0,0,0)$. This theory is somewhat singular and we shall refrain from discussing it for now. We can also close the puncture with the baryon made from $\tilde{Q}_1$. This will lead to a torus with flux $(1,0,0)$. The quiver description of this theory is shown in figure \ref{Affinequiver}. One can see that it resembles an affine $A_1$ quiver with additional singlets coupled through a superpotential. We shall refer to this theory as the affine quiver. 

\paragraph{The ``affine quiver'':}

Next we can study some of its properties. Here we shall consider the simplest though somewhat special case of $N=2$, and delay discussion of the higher $N$ cases to the next subsection. We begin with studying its anomalies. In this case the full superpotential, including the contribution from the flipping, is cubic, and performing a-maximization we indeed find that under the correct $\fu(1)_R$ all chiral multiplets have the free field $R$-charge $\frac{2}{3}$. Further we find that:
\be
a = \frac{5}{3}, \qquad \;\,c = \frac{11}{6} \label{anomAQ}  
\ee  

Note that these are the anomalies for the affine $A_1$ quiver in addition to two free hypermultiplets\footnote{These are presumably the two chiral fields that accompany the adjoints plus the two chiral fields that are introduced for the flipping.}. This fact and the values for the $R$-charges suggest that this theory has a subspace on its conformal manifold where it is indeed the affine $A_1$ quiver gauge theory, where the supersymmetry enhances to $\mathcal{N}=2$.

We can also evaluate the index of this theory. Particularly we consider the affine quiver without the singlets as these are just free fields. The subgroup of $\fso(7)$ which commutes with the flux is $\fu(1) \times \fsu(2)^2$ and so this is the expected global symmetry. Since without the singlets the matter content of this theory is the same as the $\mathcal{N}=2$ affine $A_1$ quiver, with global symmetry $\fu(1)\times \frak{usp}(4)$, the index in fact forms into characters of this group. We find it is given by:
\bea
I^{\text{Affine Quiver}}_{N=2, k=2} 
& = & 1 + p^{\frac{2}{3}} q^{\frac{2}{3}} (2 \beta^4 + \frac{1}{\beta^2} \chi[\bold{10}]_{\frak{usp}(4)}) - 2 \beta^2 (p^{\frac{4}{3}} q^{\frac{1}{3}} + p^{\frac{1}{3}} q^{\frac{4}{3}}) 
\nonumber \\
& &+ p q (1 + \chi[\bold{5}]_{\frak{usp}(4)} - \chi[\bold{10}]_{\frak{usp}(4)}) + \ldots ,
\label{AQNtwoIn}
\eea 
where $\chi[\bold{4}]_{\frak{usp}(4)} = t \gamma + \frac{1}{t\gamma} + \frac{t}{\gamma} + \frac{\gamma}{t} = [1,0]_{\frak{usp}(4)}$.

The Higgs branch of this $\mathcal{N}=2$ Affine $A_1$ theory was evaluated in \cite{Hanany:2010qu}, see section 5.2.1 and in particular equation (5.19). Furthermore it was found to be the closure of the next to minimal orbit of $\frak{usp}(4)$ as in Table 3 of \cite{Cremonesi:2014uva} and Tables 10 and 12 of \cite{Hanany:2016gbz}, where another description sets it as the $\bZ_2$ orbifold of the closure of the minimal nilpotent orbit of $\mathrm{SL}(4)$ (alternatively known as the reduced moduli space of 1 $\SU(4)$ instanton on $\bC^2$). This emphasizes that the global symmetry on this part is indeed $\frak{usp}(4)$. The unrefined Hilbert Series takes the form
\be
H^{\text{Affine Quiver}}_{N=2, k=2} (\tau) = \frac{(1+\tau^2)(1+3\tau^2+\tau^4)}{(1-\tau^2)^6}
\ee
and it admits the highest weight generating function \cite{Hanany:2014dia,Hanany:2016gbz}
\be
HWG^{\text{Affine Quiver}}_{N=2, k=2} (\tau, \mu_1, \mu_2)= \frac{1}{(1-\mu_1^2 \tau^2)(1-\mu_2^2\tau^4)},
\ee
with $\mu_1$ and $\mu_2$ the fugacities for the highest weights of $\frak{usp}(4)$. From this one deduces the refined Hilbert series that admits a character expansion
\be
H^{\text{Affine Quiver}}_{N=2, k=2} (\tau, t, \gamma) = \sum_{n_1=0}^\infty\sum_{n_2=0}^\infty [2n_1, 2n_2]_{\frak{usp}(4)} \tau^{2n_1+4n_2}.
\ee
In addition there are the $4$ singlets, which in this construction, two are given the charge $\beta^2$ and two the charge $\frac{1}{\beta^4}$. Therefore their presence does not interfere with the global symmetry. Again it is reasonable to expect that at some sub-locus of the conformal manifold these are indeed free fields and so can be rotated separately leading to additional enhancement of symmetry. We also note that this structure is common in the so called ``ugly" class ${\cal S}$ theories where the SCFT is accompanied by additional free hypers whose global symmetry is identified with part of the global symmetry of the SCFT.   

Next we can study the dimension of the conformal manifold for this theory. As the singlets cannot add additional directions \cite{Green:2010da}, it can be directly read from the index without them (\ref{AQNtwoIn}). Particularly, we look at the $p q$ order terms, which according to a result by \cite{Beem:2012yn}, are just the marginal operators minus the conserved currents. For the case at hand we find that there are $7$ marginal operators where one is canceled against the $\fu(1)$ conserved current in (\ref{AQNtwoIn}). Applying the logic of \cite{Green:2010da} we find a dimension $3$ conformal manifold, reproducing the result in section 3.2 of \cite{Benvenuti:2005wi}, along which the symmetry is broken to $\fu(1)\times \fsu(2)^2$. This is smaller than the $2k=4$ we expect from 6d arguments. This case is thus somewhat similar to the story for $k=1$. 

It is instructive to consider this theory as an $\mathcal{N}=2$ theory which can be constructed by the compactification of the $A_{1}$ $(2,0)$ theory on a torus with two maximal punctures. Then the 6d analysis of \cite{Bah:2012dg} leads us to expect a three-dimensional conformal manifold, two directions of which preserve all the symmetries and correspond to the coupling constants of the two groups while the remaining one preserving only $\mathcal{N}=1$ supersymmetry and $\fu(1)\times \fsu(2)^2$ global symmetry. This agrees with our observation.

The class ${\cal S}$ and class ${\cal S}_2$ theories differ by the existence of the singlets. It is not difficult to see one can build marginal operators uncharged under the 6d apparent global symmetries. By the logic of \cite{Green:2010da} this should lead to exactly marginal operators. However, this fails as these operators in fact become free leading to the appearance of accidental symmetries invalidating the argument. Therefore we conclude that the 6d expectations regarding the conformal manifold are too naive, and like the anomaly analysis, can be modified due to the appearance of accidental symmetries.  

\paragraph{The ``Klebanov-Witten'' theory:}

\begin{figure}[t]
\center
\includegraphics[width=0.43\textwidth]{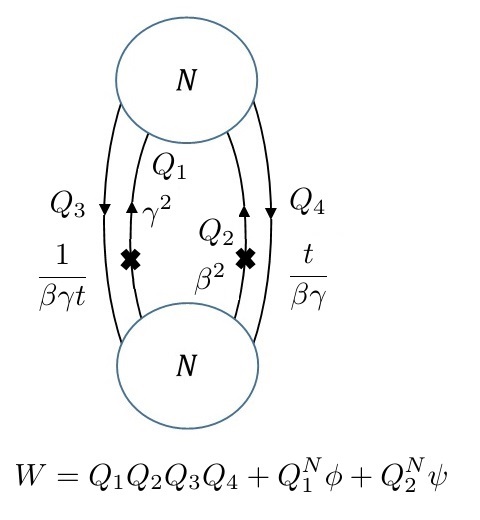} 
\caption{The quiver diagram for the 4d class $\CS_2$ theory corresponding to a torus with flux $(\frac{1}{2},\frac{1}{2},0)$. Next to the fields are their charges summarized through fugacities. The theory has a quartic superpotential involving the four bifundamentals as well as the superpotential coming from the flipping. There is also an $R$-symmetry where it is convenient to give $R$-charge $\frac{1}{2}$ to the four bifundamentals $Q_i$ and $R$-charge $2-\frac{N}{2}$ to the flipping fields $\psi, \phi$.}
\label{KWquiver}
\end{figure}

We can also close the puncture with the baryon made from $\tilde{Q}_4$. This will lead to a torus with flux $(\frac{1}{2},\frac{1}{2},0)$. The quiver description of this theory is shown in figure \ref{KWquiver}. One can see that it resembles the Klebanov-Witten model \cite{Klebanov:1998hh}, but with additional singlets $\phi, \psi,$ coupled through superpotential terms. We shall refer to this theory as the KW case.

Let us first consider the theory without the singlets. In this case this model is known to go to an interacting fixed point where the bifundamental fields have $R$-charge $\frac{1}{2}$ \cite{Klebanov:1998hh}. Now consider adding the free fields and couple them through the superpotential. The behavior of the resulting theory depends on the value of $N$. For $N>2$, this superpotental is irrelevant and the theory should flow to the same fixed point, but with free singlets. However, for $N=2$, this superpotental is relevant and the theory should flow to a new fixed point. We shall now discuss the latter case in more detail. Note that for this special case of $N=2$, $Q_1$ is a $2\times2$ matrix and the notation $Q_1^2$ stands for $\det Q_1$, and similarly for $Q_2$. Each of these terms is invariant under a corresponding $\fsu(2)$ global symmetry and the global $\fsu(4)$ symmetry which is present in the absence of these terms \cite{Forcella:2008eh} is broken to $\fsu(2)\times \fsu(2)\times \fu(1)$ in the presence of these terms, where $\fu(1)$ is the baryonic symmetry which acts as $+1$ on $Q_1$ and $Q_2$ and as $-1$ on $Q_3$ and $Q_4$.

First we shall need to perform a-maximization to determine the superconformal R-symmetry. It is straightforward to see that only the baryonic symmetry $\fu(1)_{\beta} + \fu(1)_{\gamma}$ can mix with the naive $\fu(1)_R$ of the KW model. Thus, we define:
\be
\fu(1)'_R = \fu(1)_R + \alpha \frac{\fu(1)_{\beta} + \fu(1)_{\gamma}}{2}
\ee  

By performing a-maximization we find $\alpha = \frac{\sqrt{10}-3}{6} \approx 0.027$, so the $R$-charges change only slightly compared to their naive value. One can check that all gauge invariant fields are above the unitary bound so this is consistent with the theory flowing to an interacting fixed point. 

We can next evaluate the anomalies for this theory. Particularly, for the conformal anomalies we find:
\be
a = \frac{5}{6} \sqrt{\frac{5}{2}}, \qquad\, c = \frac{11}{12} \sqrt{\frac{5}{2}}
\ee

Next we evaluate the index for this theory. First we note that the subgroup of $\fso(7)$ that commutes with the flux is $\fu(1)\times \frak{usp}(4)$ so this is the expected global symmetry, enhancing the global symmetry of $\fsu(2)\times \fsu(2)\times \fu(1)$ found above. Indeed we find that the index naturally forms into characters of this symmetry, where it is given by:
\bea
I^{\text{KW}}_{N=2, k=2} & = & 1 + p^{\frac{1}{2}} q^{\frac{1}{2}} ( \beta^2 \gamma^2 + \frac{1}{\beta^2 \gamma^2} \chi[\bold{5}]_{\frak{usp}(4)} + \chi[\bold{4}]_{\frak{usp}(4)}) + \frac{1}{\beta^2 \gamma^2}(p^{\frac{1}{2}} q^{\frac{3}{2}} + p^{\frac{3}{2}} q^{\frac{1}{2}})(\chi[\bold{5}]_{\frak{usp}(4)} - 1) \nonumber \\ & + & p q (\beta^4 \gamma^4 + \chi[\bold{10}]_{\frak{usp}(4)}+ \frac{1}{\beta^2 \gamma^2} \chi[\bold{16}]_{\frak{usp}(4)} + \frac{1}{\beta^4 \gamma^4} (1+\chi[\bold{14}]_{\frak{usp}(4)})) + ... \label{IndKW}
\eea 
where $\chi[\bold{4}]_{\frak{usp}(4)} = \frac{t \gamma}{\beta} + \frac{\beta}{t\gamma} + \frac{t \beta}{\gamma} + \frac{\gamma}{t \beta}$. Note that in (\ref{IndKW}) we have used the naive $R$-symmetry so the true $R$-charges of various operators should be shifted depending on their $\beta \gamma$ charges.

It is interesting that the singlets are necessary to get the enhanced global symmetry that is required from matching to 6d. They first break $\fsu(4)$ to $\fsu(2)\times \fsu(2)\times \fu(1)$ and then enhance to $\fu(1)\times \frak{usp}(4)$ which is evident from the index but not from the superpotential. This enhancement is only present for $N=2$ which is consistent with the superpotential being relevant only for $N=2$.

Another interesting computation is to check the moduli space for the conifold theory at $N=2$. We will use for convenience the map $\tau^2 = \sqrt{p q}$. The theory has a global symmetry $\fsu(4)$ and the Hilbert series was shown in \cite{Forcella:2008eh} and particularly in equation 4.16 to admit a character expansion of the form
\be
H\left(\tau,t, \frac{\beta}{\gamma}, \beta \gamma\right)^{\text{Conifold}}_{N=2} = \sum_{n_1=0}^\infty \sum_{n_2=0}^\infty [2n_1,2n_2,0]_{su(4)}\tau^{2n_1+4n_2},
\ee
or alternatively, using the fugacities for the highest weights $\mu_1, \mu_2$, a highest weight generating function
\be
HWG\left(\tau,\mu_1,\mu_2\right)^{\text{Conifold}}_{N=2} = \frac{1}{(1-\mu_1^2 \tau^2)(1-\mu_2^2 \tau^4)}.
\ee
These computations lead to a description of the moduli space as the set of all 4 by 4 complex symmetric matrices with rank at most 2.
The natural guess after flipping and symmetry enhancement, with highest weight fugacities $\nu_1$ and $\nu_2$ for $\frak{usp}(4)$, is given by the highest weight generating function
\be
HWG\left(\tau,\nu_1,\nu_2, \beta \gamma \right)^{\text{KW}}_{N=2} = \frac{1}{(1- \beta^2 \gamma^2 \tau^2)(1-\nu_1 \tau^2)(1-\frac{1}{\beta^2 \gamma^2} \nu_2 \tau^2)},
\ee
which leads to an unrefined Hilbert series
\be
H\left(\tau \right)^{\text{KW}}_{N=2} = \frac{1+3\tau^2 +\tau^4}{(1-\tau^2)^7}.
\ee

We can also calculate the dimension of the conformal manifold from the index. It is again given by the $p q$ order terms under the true $R$-symmetry. For our case this translates to the $p q$ order  operators which are uncharged under $\fu(1)_{\beta \gamma}$. Thus we find an $11$ dimensional conformal manifold along which the $\frak{usp}(4)$ group is completely broken. This is greater than the $4$ dimensional one we expect from 6d reasoning. This does not contradict the 6d reasoning since there could be 4d marginal operators with no clear interpretation in 6d. It is somewhat similar to the $k=1, N>2$ case.  It is also interesting to compare this with the 5 dimensional conformal manifold for the conifold theory which was found in section 2.2 of \cite{Benvenuti:2005wi}.

Another interesting observation regarding the index (\ref{IndKW}) is the appearance of the $\bold{4}$ dimensional representation of $\frak{usp}(4)$. This implies that the global symmetry is $\USp(4)$ and not $\SO(5)$. This group in 6d comes from breaking the $\SO(7)$ global symmetry of the 6d SCFT. Naively this suggests that this group must be $\Spin(7)$ and not $\SO(7)$. However, \cite{Razamat:2016dpl} found various 4d theories, matching 6d compactifications with fluxes that are consistent only with $\SO(7)$. These two observations suggest one of two scenarios. One, the 6d group is in fact $\Spin(7)$ which naturally explains the appearance of the $\bold{4}$ in the index (\ref{IndKW}). Then the 4d theories with non-standard quantization should be viewed as 4d theories with no valid 6d origin. An alternative explanation is that the 6d group is $\SO(7)$, which naturally fit the observations of \cite{Razamat:2016dpl}. However in that case one must view the spinors in the 4d index (\ref{IndKW}) as 4d operators without a 6d origin, similarly to the excess marginal operators we seem to find for this theory. This interpretation then implies that the 4d theory in fact undergoes an accidental discrete enhancement of symmetry $\SO(5) \rightarrow \USp(4)$.

\subsection{$N=3, k=2$ }

It is straightforward to also consider other values of $N$. The case of $N=2$ has some special features that are not present in the general case. We shall now discuss the behavior for general $N$ using the example of $N=3$ where actual values are needed, like in index calculations. 

\paragraph{The ``affine quiver'' theory:}
We start with the affine $A_1$ quiver case. The matter content and charges are as in figure \ref{Affinequiver}. The models contain four singlet fields, two off which come from the flipping and are coupled through a superpotential. Without this superpotential, all fields have free R-charges and the theory is expected to sit on the conformal manifold of the $\mathcal{N}=2$ affine quiver with two free hypers. The superpotential involving the flipped fields is irrelevant and so the theory with these terms is expected to flow back to the $\mathcal{N}=2$ affine quiver with singlets. Therefore, at the fixed point, there will be additional symmetries rotating the free fields.

We can now consider the index of this theory for the case of $N=3$. The $6d$ global symmetry for $k=2, N>2$ is $\fsu(2)_t \times \fsu(2)_{\gamma}\times \fsu(2)_{\beta}$. This should be broken by the flux to $\fsu(2)_t \times \fsu(2)_{\gamma}\times \fu(1)_{\beta}$ which is the symmetry we expect in the 4d theory. We indeed find that the index forms characters of that symmetry. Ignoring the singlets, as these are just free fields, we find the index to be:
\bea
I^{\text{Affine Quiver}}_{N=3, k=2} & = & 1 + p^{\frac{2}{3}} q^{\frac{2}{3}} (2 \beta^4 + \frac{1}{\beta^2} (1+\chi[\bold{3}]_{\fsu(2)_t})) - 2 \beta^2 (p^{\frac{4}{3}} q^{\frac{1}{3}} + p^{\frac{1}{3}} q^{\frac{4}{3}}) \nonumber \\ & + & p q (2 \beta^6 + \frac{1}{\beta^3}\chi[\bold{4}]_{\fsu(2)_t}\chi[\bold{2}]_{\fsu(2)_{\gamma}}) + ...
\eea   
where $\chi[\bold{2}]_{\fsu(2)_t} = t + \frac{1}{t}$ and $\chi[\bold{2}]_{\fsu(2)_{\gamma}} = \gamma^3 + \frac{1}{\gamma^3}$.

Additionally there are the $4$ singlets which in this construction two are given the charge $\beta^2$ and two the charge $\frac{1}{\beta^6}$. Therefore their presence does not interfere with the symmetry.

We can also compare the dimension of the conformal manifold with the 6d expectations. Again we find that the dimension of the conformal manifold is in fact greater than what is expected from 6d.

\paragraph{The ``Klebanov-Witten'' theory:}

Next we consider the KW case. As we previously discussed for $N>2$ the superpotential coupling the singlets is irrelevant and the theory should flow to the KW model with singlets.

Next we evaluate the index for $N=3$. The flux in this case is $(\frac{1}{2},\frac{1}{2},0)$ and we expect the $4d$ theory to preserve an $\fsu(2)_t \times \fu(1)_{\frac{\gamma}{\beta}}\times \fu(1)_{\gamma\beta}$ global symmetry. In fact we shall see that the KW model shows an $\fsu(2)_t \times \fsu(2)_{\frac{\gamma}{\beta}}\times \fu(1)_{\gamma\beta}$ global symmetry which is broken due to the singlets, at least for $N=3$ and to the evaluated order. Thus, this theory may have a considerable enhancement of symmetry in the IR. As the singlets decouple in the IR we concentrate only on the interacting part, for which we find the index to be:   
\bea
I^{\text{KW}}_{N=3, k=2} & = & 1 + p^{\frac{1}{2}} q^{\frac{1}{2}} \chi[\bold{2}]_{\fsu(2)_t}\chi[\bold{2}]_{\fsu(2)_{\frac{\gamma}{\beta}}} + p^{\frac{3}{4}} q^{\frac{3}{4}} (\frac{1}{\beta^3 \gamma^3}\chi[\bold{4}]_{\fsu(2)_t} + \beta^3 \gamma^3 \chi[\bold{4}]_{SU(2)_{\frac{\gamma}{\beta}}}) \nonumber \\ & + & p q (2 + 2\chi[\bold{3}]_{\fsu(2)_t}\chi[\bold{3}]_{\fsu(2)_{\frac{\gamma}{\beta}}} - \chi[\bold{3}]_{\fsu(2)_t} - \chi[\bold{3}]_{\fsu(2)_{\frac{\gamma}{\beta}}}) + ... \label{IndKW1}
\eea 
where $\chi[\bold{2}]_{\fsu(2)_t} = t + \frac{1}{t}$ and $\chi[\bold{2}]_{\fsu(2)_{\frac{\gamma}{\beta}}} = \frac{\gamma}{\beta} + \frac{\beta}{\gamma}$.

In addition there are two free singlets with charges $\frac{1}{\gamma^6}$ and $\frac{1}{\beta^6}$. These are inconsistent with $\fsu(2)_{\frac{\gamma}{\beta}}$ implying that it is broken to its Cartan by the superpotential only to return in the deep IR. This again resembles some situations in class ${\cal S}$ theories where the global symmetry of an interacting theory plus hypers is broken by mixing part of it with the symmetry rotating the hypers.

The results of equation (\ref{IndKW1}) agree with the computations in equation 3.83 of \cite{Forcella:2007wk} but are still missing two essential operators that transform as $\chi[\bold{2}]_{\fsu(2)_t}\chi[\bold{1}]_{\fsu(2)_{\frac{\gamma}{\beta}}}$ and $\chi[\bold{1}]_{\fsu(2)_t}\chi[\bold{2}]_{\fsu(2)_{\frac{\gamma}{\beta}}}$ -- the so called non factorizable baryons. It will be interesting to check if higher order computations produce these two essential contributions.

We can also compare the dimension of the conformal manifold with the 6d expectations. Again we find that the dimension of the conformal manifold is in fact greater than what is expected from 6d.

We can in principle look at higher values of $N$ and even the large $N$ behavior. In fact both of the theories considered here, without the singlets, have well known large $N$ gravity duals. As the number of singlets is order $1$, it is reasonable that most of the properties of these theories will be well described by the gravity duals. In this regard it is interesting that both theories are reached by the compactification of the same 6d SCFT on the same surface differing by order $1$ fluxes.   The models we consider here correspond to having vanishing flux for $\fu(1)_t$ as in this case we have known Lagrangians. With the flux for 
$\fu(1)_t$    the models are expected to be strongly coupled, see appendix. The properties of these two types of models are qualitatively different. For example, the anomalies scale as $N^2$ for the gauge theories we consider here (when the singlet fields are appropriately taken into account), and are expected to scale as $N^3$ for the strongly coupled types as can be inferred from the six dimensional analysis. As from the six dimensional perspective we cannot infer existence of accidental symmetries this is just an expectation which can be invalidated in several ways.

\subsection{$ N=2, k=3$, The orbifold $\bC^3/\bZ_2\times \bZ_3$}

Finally we consider a case with $k>2$ specifically $k=3$. We shall keep $N$ general in the derivation, but for particular examples we shall use $N=2$. To derive the theories we consider taking three free trinions, connecting them together and closing three minimal punctures. More specifically we shall close two punctures with a vev to baryons charged under $\beta_1$ with the same charge.

\begin{figure}[t]
\center
\centerline{\includegraphics[width=.51\textwidth]{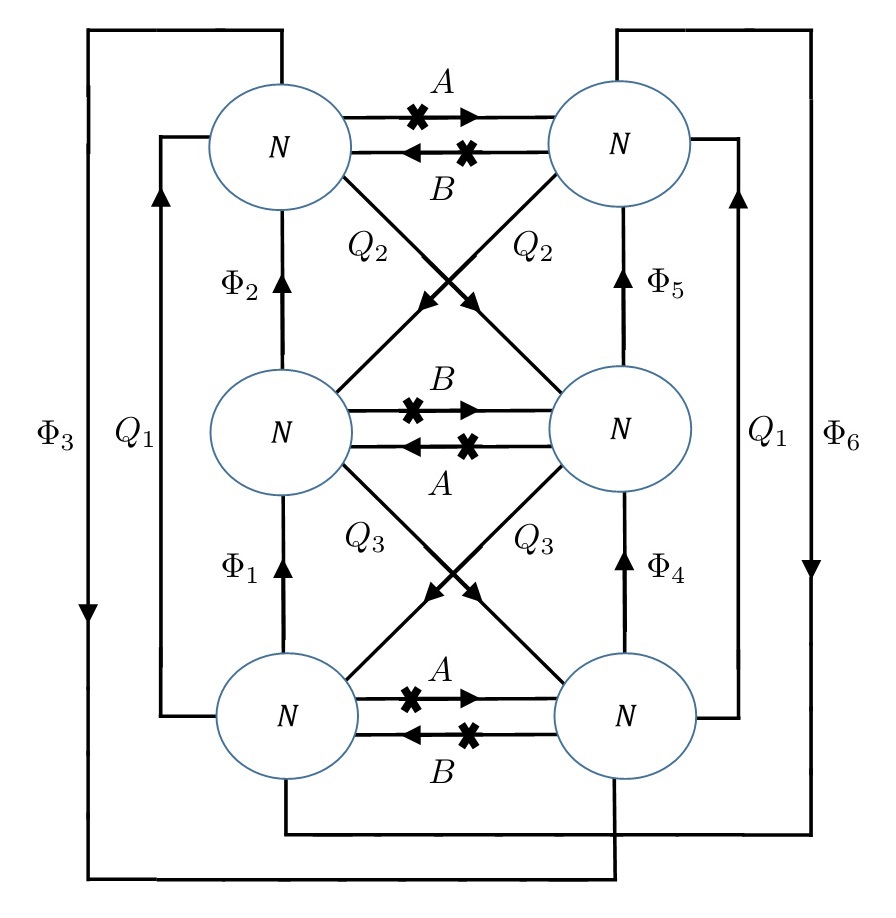}\;\;\;\;\;\;\;\;\;\,\,\,\,\,\,\, \includegraphics[scale=1.3]{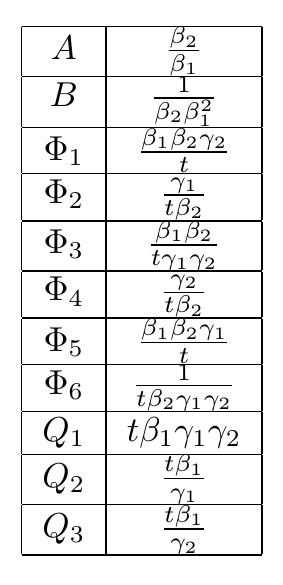} } 
\caption{On the left is the quiver diagram for the 4d class $\CS_3$ theory corresponding to a torus with only $\beta_1$ flux, while on the right is a table summarizing the charges of the various fields. Note that several different fields have the same charges and so are represented with the same letter. This theory has a rather large cubic superpotential involving the 12 triangles in the diagram. Again these are most conveniently generated by taking all cubic terms consistent with the symmetries. Additionally there are the superpotential terms coming from the flipping, which in general are not cubic. It is again convenient to choose the $R$-symmetry so that all non-flipping fields have $R$-charge $\frac{2}{3}$.}
\label{kthreequiver}
\end{figure}

Now we need to close the final minimal puncture. We consider two different possibilities. First we consider closing the last puncture also with a vev to a baryon charged under $\beta_1$ with the same charge as the last two. This is similar to how we got the affine quiver in the $k=2$ case and leads to the theory shown in figure \ref{kthreequiver}. 

In this theory all gauge groups see $3N$ flavors and so are conformal. Thus, without the flipping superpotential, all field have the free $R$-charge $\frac{2}{3}$. The flipping superpotential is marginal for $N=2$ and irrelevant for $N>2$ so either way all fields should have $R$-charge $\frac{2}{3}$ under the superconformal $R$-symmetry. Again for $N>2$ this entail an IR enhancement of symmetry due to the flipping fields becoming free.

We can consider the index for the case of $N=2$. Again we shall concentrate on the interacting part and ignore the singlets as these are free. The 6d global symmetry here is enhanced to $\fsu(6)$, but our choice of flux breaks it to $\fsu(3)_{\gamma} \times \fsu(2)\times \fu(1)_t \times \fu(1)$. Evaluating the index, we find that it can be written as:
\bea
I^{\text{All } \beta_1}_{N=2, k=3} & = & 1 + p^{\frac{2}{3}} q^{\frac{2}{3}} ( \frac{3}{\beta^3_1}(1+\chi[\bold{2}]_{\fsu(2)}) + 2 t^2 \beta^2_1 \chi[\bold{\bar{3}}]_{\fsu(3)_{\gamma}} + \frac{\beta_1}{t^2} \chi[\bold{2}]_{\fsu(2)}\chi[\bold{3}]_{\fsu(3)_{\gamma}}) \nonumber \\ & + & p q (3\chi[\bold{2}]_{\fsu(2)}  + 2 t^3\beta^3_1 + 2\frac{\beta^2_1}{t}\chi[\bold{\bar{3}}]_{\fsu(3)_{\gamma}}) + ... \label{IkthreeAf}
\eea   
where $\chi[\bold{2}]_{\fsu(2)} = \beta^2_2 \beta_1 + \frac{1}{\beta^2_2 \beta_1}$ and  $\chi[\bold{3}]_{\fsu(3)_{\gamma}} = \gamma^2_1 + \gamma^2_2 + \frac{1}{\gamma^2_1\gamma^2_2}$. Additionally there are $6$ singlet fields, $3$ with charge $\frac{\beta^2_1}{\beta^2_2}$ and $3$ with charge $\beta^4_1 \beta^2_2$. These can be written as $3 \beta^3_1 \chi[\bold{2}]_{\fsu(2)}$ and so are consistent with the $6d$ global symmetry. 

We can also consider the dimension of the conformal manifold. By the reasoning of \cite{Green:2010da}, of the terms appearing in (\ref{IkthreeAf}), the ones charged under $\beta_1$ do not contribute any exactly marginal operators as one cannot form a $\beta_1$ invariant from them. This leaves the three marginal operators in the $\chi[\bold{2}]_{\fsu(2)}$, as well as the marginal operators in the adjoint of the full global symmetry $G$ which must be present to cancel the contribution of the conserved currents. These contribute $10$ exactly marginal operators. So we now have more than those expected from (\ref{ConfM}).

\begin{figure}[t]
\center
\includegraphics[width=0.57\textwidth]{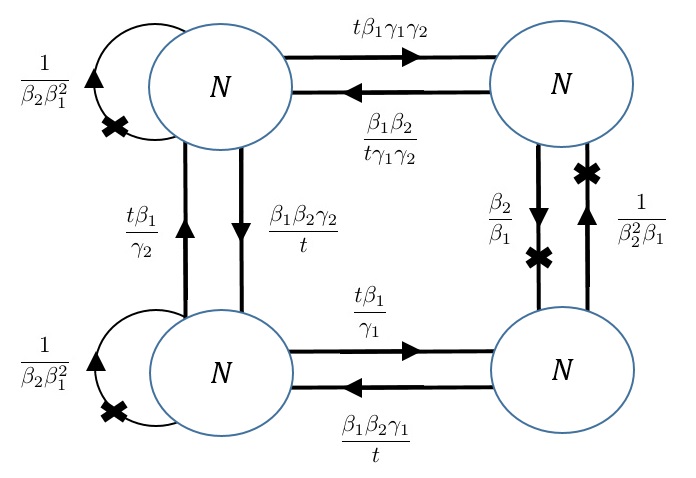} 
\caption{The quiver diagram for the 4d class $\CS_3$ theory corresponding to a torus with two units of $\beta_1$ flux and one of $\beta_2$ flux. Next to the fields are their charged summarized using fugacities. This theory has a combination of cubic and quartic superpotential terms. Again these are most conveniently generated by taking all terms consistent with the symmetries. Additionally there are superpotential terms coming from the flipping. The theory also has an $R$-symmetry, a convenient choice for which is to give all the bifundamentals $R$-charge $\frac{1}{2}$, and R-charge $1$ for the adjoints and their singlets.}
\label{Mixedquiver}
\end{figure}

We can also consider closing the last puncture with a baryon charged under $\beta_2$. The resulting flux leads to the symmetry breaking pattern $\fsu(6)\rightarrow \fsu(4)\times \fsu(2)\times \fu(1)$ which we expect to be the 4d global symmetry. In the field theory the vev leads to a quiver with an $\fsu(N)$ group with $N$ flavors. This group confines in the IR leading to the identification of the groups it's connected to and making the flipping fields massive. After the dust settles we end with the so called $L^{222}$ \cite{Franco:2005sm} quiver theory in figure \ref{Mixedquiver}. This theory can also be derived from 4 NS branes on the circle, with two types of orientation.

We next proceed to analyze this theory in detail. First we need to evaluate the conformal R-symmetry. By inspection one can see that there is only one $\fu(1)$ that can mix with the R-symmetry, which in our notation is $\fu(1)_{\beta_1}$. The remaining $\fu(1)$'s can be grouped into $4$ baryonic $\fu(1)$'s, each rotating one of the four pairs of bifundamentals with opposite charges while the adjoints and their associated singlets being neutral. Three of these $\fu(1)$'s are combinations of $\fu(1)_t, \fu(1)_{\gamma_1}$ and $\fu(1)_{\gamma_2}$ while the last is $2\fu(1)_{\beta_2} - \fu(1)_{\beta_1}$. 

Next we preform a-maximization. We take the $R$-symmetry to be:
\be
\fu(1)'_R = \fu(1)_R + \alpha \fu(1)_{\beta_1}
\ee
where we take $\fu(1)_R$ to rotate the bifundamentals with charge $\frac{1}{2}$ and the adjoints and their associated singlets with charge $1$. The flipping fields attached to the bifundamental then also have R charge $1$ while those attached to the adjoits have R charge $0$. Performing the a-maximization we find that: $\alpha=\frac{3-\sqrt{5}}{6}$. With this value we find that $a = \frac{5\sqrt{5}}{4}$.

 However with this R-charge the adjoint flipping fields are below the unitary bound. Therefore the natural conjecture is that these fields become free at some point along the flow leading to an accidental $\fu(1)$ that mixes with the R-symmetry. We can now repeat the a-maximization, but taking these to be free fields where we find: $\alpha=\frac{\sqrt{13}-3}{6}$. We find that all fields have dimensions above the unitary bound and that the superpotential coupling the flipping fields to the adjoints is irrelevant. All of these are consistent with our claim. We can also preform a-maximization considering all the flipping fields as free, where we indeed find that, compared to that point, the superpotential coupling the flipping fields to the adjoints is irrelevant while the one coupling the flipping fields to the bifundamentals is relevant.

So to conclude we expect the theory in figure \ref{Mixedquiver} to flow to an interacting fixed point consisting of the quiver theory, without the adjoint flipping, plus two free chiral fields. We next want to evaluate the index of this fixed point. Again for simplicity we shall first ignore the two free chiral fields. From 6d we expect an $\fsu(4)\times \fsu(2)\times \fu(1)$ global symmetry. We indeed find that the index can be written in characters of this symmetry, where it reads:
\bea
I^{\text{Mixed}}_{N=2, k=3} & = & 1 + p^{\frac{1}{2}} q^{\frac{1}{2}} ( \frac{3}{\beta^2_1 \beta_2} + \beta^2_1 \beta_2 \chi[\bold{6}]_{\fsu(4)} + \beta^2_1 \beta_2 \chi[\bold{3}]_{\fsu(2)} )  \nonumber \\
&+&  \beta^2_1 \beta_2 (p^{\frac{3}{2}} q^{\frac{1}{2}} + p^{\frac{1}{2}} q^{\frac{3}{2}}) (\chi[\bold{3}]_{\fsu(2)} + \chi[\bold{6}]_{\fsu(4)} - 1) \nonumber \\ & + & p q (\frac{8}{\beta^4_1 \beta^2_2} + 1 +  \beta^4_1 \beta^2_2  \chi[\bold{5}]_{\fsu(2)} + \beta^4_1 \beta^2_2  \chi[\bold{3}]_{\fsu(2)}\chi[\bold{6}]_{\fsu(4)} + \beta^4_1 \beta^2_2 \chi[\bold{20'}]_{\fsu(4)} \nonumber \\ & + & \beta^2_1 \beta_2 \chi[\bold{2}]_{\fsu(2)}( \chi[\bold{4}]_{\fsu(4)} + \chi[\bold{\bar{4}}]_{\fsu(4)}) + 2\chi[\bold{6}]_{\fsu(4)} + 2\chi[\bold{3}]_{\fsu(2)}) + ...
\eea   
where $\chi[\bold{2}]_{\fsu(2)} = \beta^{\frac{3}{2}}_2 + \frac{1}{\beta^{\frac{3}{2}}_2}$ and $\chi[\bold{4}]_{\fsu(4)} = \frac{t}{\sqrt{\beta_2}} (\gamma^2_1 + \gamma^2_2 + \frac{1}{\gamma^2_1\gamma^2_2}) + \frac{\sqrt{\beta^3_2}}{t^3}$. Note that here we have used the naive $R$-symmetry, not the superconformal one, so the dimensions of operators are shifted based on their $\beta_1$ charge.

One can see that, as expected, the $\fu(1)$ is given by $\fu(1)_{\beta_1}$ while $2\fu(1)_{\beta_2} - \fu(1)_{\beta_1}$ forms the non-abelian part. This is apparent as the fugacity of the $\fu(1)$ is $\beta^2_1 \beta_2$ so states charged only under it are invariant under a $2\fu(1)_{\beta_2} - \fu(1)_{\beta_1}$ transformation. Alternatively those charged only under the non-abelian part are invariant under $\fu(1)_{\beta_1}$ transformations. 

Finally we note that the dimension of the conformal manifold is again greater than what is expected from 6d.

\

\section{Compactification with discrete twists}
\label{sec:global}

For now we were satisfied with the discussion of the symmetries of the models at the level of the algebra rather than the group. However, the global properties of the group in six dimensions have far reaching implications as far as the choice of fluxes goes. As we will next discover the six dimensional constructions have a non trivial global structure which allows for discrete fluxes to be switched on which in particular break some of the continuous global symmetry. This procedure has a four dimensional analogue. We turn our attention to this next.

\subsection{6d analysis}
\paragraph{Global form of the flavor symmetry:}
There is no doubt that the 6d \Nequals{(1,0)} theory we have been using in this paper has the flavor symmetry $\fsu(k)\times\fsu(k)\times \fu(1)_t$ at the Lie algebra level. 
What is exactly the flavor symmetry group? 
Let us for now concentrate our attention to the subgroup connected to the identity, neglecting the $\fu(1)_t$ part.
On the generic point of the tensor branch, the theory becomes a linear quiver gauge theory.
It contains a gauge-invariant operator which is a bifundamental of $\fsu(k)\times\fsu(k)$, obtained by multiplying all the bifundamentals of the quiver.
This suggests that the group is \begin{equation}
\frac{\SU(k)\times \SU(k)}{\bZ_k}
\end{equation}  where the quotient is with respect to the diagonal combination of the two centers.

If this is the case, on a compactification on $T^2$, we should be able to turn on the (generalized) Stiefel-Whitney class $w_2\in \bZ_k$.  
In the next subsection, we will find the corresponding operation in the 4d field theory language, and will check that the anomaly computed in 4d agrees with the expectation from 6d. 

Before doing that, let us remind ourselves the basics of the (generalized) Stiefel-Whitney class $w_2$ of bundles of non-simply-connected groups on $T^2$. (A detailed account readable for physicists can be found in \cite{Witten:1999ds,Witten:2000nv}. For complete generality, the reader should consult \cite{Borel:1999bx}.)
Let us say we construct a $G$ bundle on $T^2$ by first having a $G$-bundle on a rectangle by identifying two sets of parallel edges. This identification involves specifying the gauge transformation used in the gluing process along the boundary of the rectangle. This becomes a closed path within the group manifold, which is topologically classified by $\pi_1(G)$.
From this we see that when $G=\SU(k)/\bZ_k$ the bundle is classified by a number in $\bZ_k$.
This is the (generalized) Stiefel-Whitney class $w_2$.
When the Stiefel-Whitney class of an $\SU(k)/\bZ_k$ bundle is nontrivial, we cannot think of it as an $\SU(k)$ bundle. 
Next we discuss  two easy ways to construct an $\SU(k)/\bZ_k$ bundle with a nontrivial Stiefel-Whitney class. 

\paragraph{Flat bundles with nontrivial $w_2$:}
First are the flat bundles. In this case we have the holonomy $g_A$ along the $A$ cycle and the holonomy $g_B$ along the $B$ cycle. They should commute in $\SU(k)/\bZ_k$. In other words, they should commute up to the center in $\SU(k)$. We can easily find such a pair: take \begin{equation}
g_A=\diag(1,\omega,\omega^2,\ldots,\omega^{k-1}), \qquad \omega=e^{2\pi i/k}\label{eq:ga}
\end{equation}
and
\begin{equation}
g_B=\begin{pmatrix}
0 & 1 \\
&0&1 \\
&&0&1 \\
&&&\ddots \\
1&&&&0
\end{pmatrix}.\label{eq:gb}
\end{equation}
Then we see $g_Ag_B=\omega g_B g_A$. This corresponds to the Stiefel-Whitney class $1\in \bZ_k$, and 
for physicists these matrices were familiar from the work of 't Hooft \cite{'tHooft:1979uj}.
There is no unbroken symmetry.

For $\ell\in \bZ_k$ such that $\gcd(\ell,k)= m$, one can easily construct $g_A,g_B$ such that $g_Ag_B=\omega^\ell g_B g_A$ such that the unbroken symmetry is $\fsu(m)$.
This is done by firstly recalling that $\SU(m)\times \SU(n) \subset \SU(k=mn)$, secondly taking $g_A$, $g_B$ in $\SU(n)$ with $\ell/m \in \bZ_n$, and regarding the resulting $g_{A,B}$ as matrices in $\SU(k)$. As the $\SU(m)$ part is untouched, clearly the flavor symmetry is $\fsu(m)$.

\paragraph{Abelian bundles with nontrivial $w_2$:}
Another are Abelian bundles. Let us denote by $T$ the Cartan torus of $\SU(k)$. Then the Cartan torus of $\SU(k)/\bZ_k$ is $T/\bZ_k$. 

An Abelian $\SU(k)$ bundle is specified by a map $\U(1) \to T$. In particular, specifying one on $T^2$ corresponds to specifying a point in a lattice $\Lambda$ of rank $k-1$ that can be naturally identified with the root lattice of $\SU(k)$. They can be thought of as $k$ integers summing to zero.

An Abelian $\SU(k)/\bZ_k$ bundle is now specified by a map to $T/\bZ_k$. Again it is specified by a point in a lattice $\Lambda^*$ again of rank $k-1$, but now identified with the weight lattice of $\SU(k)$. If we use the Chern classes normalized to the subgroup of $\SU(k)$, they  can now look  rational, with denominator $k$. The Stiefel-Whitney class can be easily read off by considering $\Lambda^*/\Lambda \simeq \bZ_k$.

\paragraph{6d configuration:}
So far we talked about $\SU(k)/\bZ_k$ bundles, but in fact we need to consider $(\SU(k)\times \SU(k))/\bZ_k$ bundles. This means that two $\fsu(k)$ parts have the same Stiefel-Whitney class. 
For example, one can choose flat bundles for both, Abelian bundles for both, or a flat bundle for one and an Abelian bundle for the other. 
The computation of the 4d anomaly is straightforward: one just has to plug in the Chern classes of the Abelian parts in the formulas we have been using.

\paragraph{Stiefel-Whitney class and the symmetry of the quiver graph:}
Now, to bridge our discussion here to the 4d analysis below, consider first the compactification to 5d. Let us put the $\SU(k)$ holonomy $g_A$ \eqref{eq:ga} around $\mathbb S^1$. Then we have a circular $\SU(N)^k$ quiver with the same gauge coupling for all groups. 
Now, the operation $g_B$ \eqref{eq:gb} naturally corresponds to the symmetry of the circular quiver shifting the node by one. 
Therefore, by compactifying the 5d theory on an additional $\mathbb S^1$ with a twist rotating the circular quiver, we can realize the compactification of the 6d theory with a nontrivial Stiefel-Whitney class.

\subsection{4d analysis}

We have constructed tori theories by combining free trinions in multiples of $k$. This way we always glued punctures of the same color and preserve all the internal symmetries. It is however possible to glue punctures of different colors at the price of breaking some of the internal symmetry. For example, constructing a loop out of $l$ free trinions the group is $\fsu(k)_\gamma \times \fu(1)_t \times {\fsu(\gcd({k,l}))}_\beta$. We can then close the punctures and try to identify the six dimensional compactification leading to such theories.

\paragraph{Gluing two punctures of a single trinion:}
Let us consider taking a free trinion and gluing the two maximal punctures to each other. The symmetry preserved here is $\fsu(k)_\gamma \times u(1)_t$ and the puncture symmetry $\fu(1)_\delta$. The theory is the affine quiver with $k$ nodes and with a singlet field associated to every node coupling to charged fields the same manner as the adjoint chirals. The theory is superconformal with the coupling of the singlet fields being marginally irrelevant leading to them decoupling in the IR as free fields and the symmetry enhancing in the IR to $\fsu(k)_\gamma \times \fu(1)_t \times \fu(1)_\delta \times \fu(k)$ with the last factor rotating the free fields. 
The adjoint chiral fields are charged $t^{\frac12}\delta^{-1}$, and the bifundmental fields between $i-1$ and $i$ node have charges $t^{\frac12}\gamma_i^{-1} \delta$ and $\frac{pq}t \gamma_i$. We can close the minimal puncture giving a vacuum expectation value to one of the baryons weighed $t^{\frac{N}2} \gamma_j^{-N}\delta^N$. Closure of the minimal puncture also entails flipping the rest of the baryons charged under $\gamma_{i\neq j}$. The model in the IR is the affine quiver with $k-1$ nodes and with singlets. The bifundamental chirals are charged $\gamma_i^{-1}\gamma_j$, $\frac{p q}t \gamma_{i}$. The singlets are $k-1$ having charges $t\; \gamma_j^{-1}$ as the chiral adjoint fields, and $k-1$ chiral fields which flip the baryons and have charges $p q \gamma_i^{N}\gamma_j^{-N}$. The chiral fields couple through either irrelevant or marginally irrelevant terms and thus flow to free fields in the IR. The theory in the IR is then a collection of $2k-2$ free chiral fields and the affine quiver with $k-1$ nodes.

\begin{figure}[h]
\center
\includegraphics[width=0.65\textwidth]{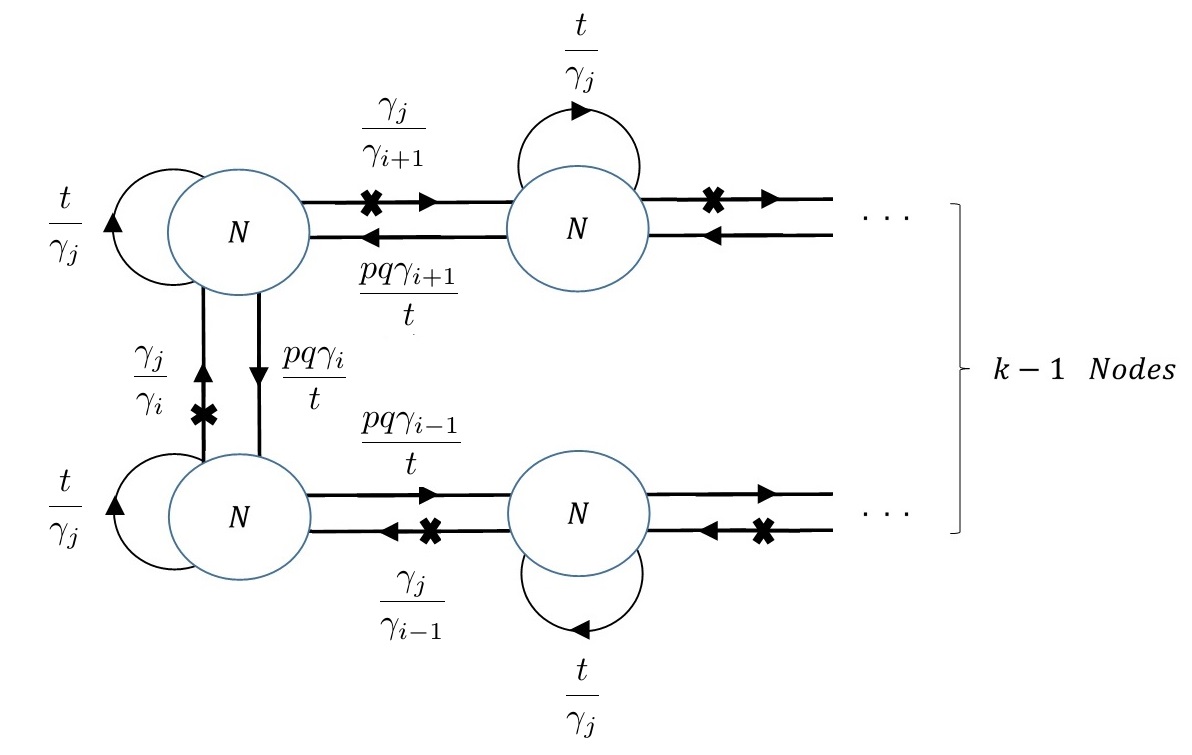} 
\caption{Toric quiver which is associated to compactification on torus with minimal punctures. }
\label{quive}
\end{figure}

The anomalies of this model match the anomalies of six dimensional compactification with flux $\frac{k-1}k$ for $u(1)_j$ and fluxes $-\frac1k$ for the $u(1)_i$ with $i$ not equal to $j$. Such fluxes are consistent if the group is $\SU(k)/{{\mathbb{Z}}_k}$ and is consistent with the general rules of association of fluxes to flows. This picture also predicts that the affine quiver should have loci on conformal manifold with the flavor group being $\fsu(k-1)_\gamma \times \fu(1)_{\gamma_j}\times \fu(1)_t$. This is not trivial from the Lagrangian of the theory which for general number of branes exhibits $\fu(1)^k$ symmetry. The index will be consistent with this claim because the affine quiver has dualities interchanging minimal punctures. This implies that the index will be invariant for permutations of $\gamma_i$. This permutation symmetry implies that the index can be written in terms of characters of $\fsu(k-1) \fu(1)_{\gamma_j}$ parametrized by $\gamma_i$. Moreover at order $pq$ of the index (when one takes the free superconformal R charges to chiral fields) the index has only singlets as the only marginal operators and conserved currents at the ${\cal N}=2$ loci of the conformal manifold are such. Thus, we expect the index to be consistent with the enhanced symmetry.

\paragraph{Gluing maximal punctures  with a shift:}

Let us consider another operation we can legally do on the field theory side. Gluing together two maximal punctures of a given theory we can twist them relatively to each other. The twist corresponds to matching $\beta_i$ to $\beta_{i+u}$ and 
$\gamma_i$ with $\gamma_{u+i}$. This breaks the global symmetry to $\fsu(\gcd(k,u))\times \fsu(\gcd(k,u))\times \fu(1)
$. These operations are considered for toric quivers in \cite{Hanany:1998it}.
 Performing both operation of gluing punctures with color differing by $l$ units and twisting with $u$  we obtain the group, $\fsu(\gcd(k,l,u))\times \fsu(\gcd(k,u))\times \fu(1)$. We can also consider gluing punctures of same orientation which will exchange roles of $\beta$ and $\gamma$ symmetries.

For the anomaly polynomials that we have computed the effect of the twist and the gluing is simple. We only have to remove the parameters which correspond to symmetries which we have broken. Namely we need to take $s^\beta_i=s^\beta_{i+l}$ and $s^\beta_{i+u}=s^\beta_i$ and $s^\gamma_{u+i}=s^\gamma_i$. Thus we conclude that the anomaly polynomial is still given by the same expressions. This also means that the matching between the 6d and 4d analysis persists also for these cases.

\paragraph{Matching with 6d:}

We are now in a position to merge the 4d and 6d observations in this section to a consistent picture. On the 4d side we observed that we can form a torus by gluing together $l$ free trinions and closing the minimal punctures. When gluing to form the torus we may also twist the gluing by say $u$ units so that in the last gluing we match $\beta_i$ to $\beta_{i+u}$. Either of these breaks the global symmetry unless $l, n$ are a multiplet of $k$.

For simplicity let us consider each of these separately. First we have seen that if we take $l = n = 0$ mod $k$ we do not break the Cartan part of the global symmetry and the anomaly polynomial calculations in 4d and 6d agree. The matching, specifically equation (\ref{maraet}), implies that closing a minimal puncture shifts the flux in a $\U(1)$ embedded inside one of the $SU(k)$'s so that its commutant is $\SU(k-1)$ and that $\bold{k} \rightarrow \bold{1}^{k-1} + \bold{k-1}^{-1}$. In this normalization the flux is shifted by $\frac{1}{k}$, and is only consistent for $\SU(k)/{{\mathbb{Z}}_k}$. As long as $l = 0$ mod $k$ the total flux in the anti-diagonal $\SU(k)$ is integer. This is a consistent flux for $(\SU(k)\times \SU(k))/\mathbb{Z}_k$. Note that if the fluxes of both $\SU(k)$'s are integer then this describes an $\SU(k) \times \SU(k)$ consistent flux. However if the fluxes of both $\SU(k)$'s are fractional then we have a non-trivial Stiefel-Whitney class on the torus that is materialized through abelian fluxes for both $\SU(k)$'s. This in general breaks the global symmetry to it's Cartan subalgebra. 

Now let's consider the case where $l = 0$ mod $k$, but $n$ has a non trivial value. In this case the global symmetry is broken down at least to $\fsu(\gcd(k,u))\times \fsu(\gcd(k,u))\times \fu(1)$, and may be further broken to the Cartan due to the fluxes. In this case we identify $n$ with a non-trivial Stiefel-Whitney class on the torus that is manifested using flat connections for both $\SU(k)$'s. This naturally accounts for the global symmetry breaking pattern. Also it is quite natural from the 5d viewpoint. It is also consistent with the matching of the anomaly polynomial between 6d and 4d as besides the breaking of symmetries this does not effect either of the calculations. Note that again if the fluxes of both $\SU(k)$'s are fractional then we have an additional non-trivial Stiefel-Whitney class so that the total Stiefel-Whitney class, which may be trivial, is materialized partially by flat connections and partially by abelian fluxes.
  
Next we consider the case where $l$ is arbitrary while $n=0$ mod $k$. Now one $\SU(k)$ is broken down at least to $\fsu(\gcd(k,l))$. We identify $l$ with a non-trivial Stiefel-Whitney class on the torus that is manifested using flat connections for one $\SU(k)$, the broken one, and abelian fluxes for the other. This correctly accounts for the symmetry breaking pattern, and also agrees with the matching of the anomaly polynomial since the two match where we have fractional fluxes for the unbroken symmetry. Note that the flat connections are necessary so that the total configuration be consistent with $(\SU(k)\times \SU(k))/\mathbb{Z}_k$.

Finally we can consider an arbitrary configuration with any value of $l$ and $n$. This should correspond to the $6d$ theory on torus with fluxes determined by $l$ through \ref{maraet} and with an $n$ Stiefel-Whitney class manifested using flat connections. From the discussion so far it is apparent that the symmetry breaking pattern as well as the matching of the anomaly polynomial are consistent with this. Note that in the generic case the total Stiefel-Whitney class is manifested using both flat connections and abelian fluxes and may vanish even when $l, n \neq 0$ mod $k$. We also note that this discussion points out that in order to specify the 6d configuration we must enumerate the Stiefel-Whitney class in addition to the fluxes. 

\section*{Acknowledgements}
We are grateful to K. Intriligator,  H. C. Kim, Z. Komargodski, and C. Vafa for useful conversations.
The work of Y.T. is  supported in part by JSPS Grant-in-Aid for Scientific Research No. 25870159.
The work of Y.T. and G.Z. is supported by World Premier International Research Center Initiative
(WPI Initiative), MEXT, Japan.  SSR is  a Jacques Lewiner Career Advancement Chair fellow. The research of SSR was also supported by Israel Science Foundation under grant no. 1696/15 and by I-CORE  Program of the Planning and Budgeting Committee.
The Work of I.B. is supported by UC president's post-doctoral fellowship and in part by DOE grant DE-SC0009919. A.~H.~ is supported by STFC Consolidated Grant ST/J0003533/1, and by EPSRC Programme Grant EP/K034456/1.

\appendix

\section{Anomalies of interacting trinions for general $k, N$}
\label{sec:trinion}
  In section \ref{sec6d}, we have obtain the anomaly coefficients of class $\CS_k$ theories from the anomaly polynomial of the six-dimensional theory.
  By using these, we now predict the anomalies of the trinion models in this section.
  
  First of all, let us reproduce here the anomalies of the class $\CS_k$ theory associated to genus $g$ Riemann surface 
    \bea
     \Tr R'
     &=&  - \frac{(k^2-2)(N-1)}{2} (2g-2), ~~~~
     \Tr t
      =   - k^2N N_s,
             \nonumber \\
     \Tr R'^3
     &=&     \frac{(N-1)(k^2(N^2+N-1)+2)}{2} (2g-2), 
               ~~~~~
     \Tr R'^2 t
      =     \frac{k^2 N (N^2 - 1)}{3} N_s,
               \nonumber \\
     \Tr R' t^2
     &=&   - \frac{k^2 N (N^2-1)}{6} (2g -2), ~~~~~
     \Tr t^3
      =   - k^2 N^3 N_s.
      \label{genusg}
     \eea 
   We omitted the anomalies involving $\beta_i$ and $\gamma_i$ which depend on the $\fsu(k)_\beta \times \fsu(k)_\gamma$ fluxes for simplicity.  
   By subtracting the contribution of the $3g-3$ tubes from these, one would get the anomalies of $2g-2$ trinions from which one can deduce
   the single contribution.
   
   To be more explicit, let us choose the case with the flux associated to the surface $N_s = g-1$.
   There is a duality frame where the four-dimensional theory consists of $2g-2$ trinion theories, each of which has $N_s=1$
   and three maximal punctures with $\sigma=+1$, combined by $3g-3$ tubes with $N_s = 0$.
  The tube is assumed to be  the $\Phi$ gluing corresponding to the $SU(N)^k$ vector multiplets and the bifundamental chiral multiplets between them.   Thus one tube contributes to the anomalies as
     \bea
     \Tr R'
     &=&    \Tr R'^3
      =     k(N^2-1), ~~
     \Tr t
      =     \Tr t^3
      =   - kN^2, ~~~
     \Tr R'^2 t
      =     \Tr R' t^2
      =     0. \label{anomaliestube}
     \eea
   By subtracting these tube contributions from \eqref{genusg} and dividing by $(2g-2)$, we obtain the anomalies of the trinion theory
   with $N_s=\half$:
     \bea
     \Tr R'
     &=&  - \frac{(k^2-2)(N-1) + 3k(N^2-1)}{2}, ~~~~
     \Tr t
      =   - \frac{k^2N - 3 k N^2}{2} ,
             \nonumber \\
     \Tr R'^3
     &=&     \frac{(N-1)(k^2(N^2+N-1)+2) - 3 k (N^2-1)}{2}, 
               ~~~~~
     \Tr R'^2 t
      =     \frac{k^2 N (N^2 - 1)}{6},
               \nonumber \\
     \Tr R' t^2
     &=&   - \frac{k^2 N (N^2-1)}{6}, ~~~~~
     \Tr t^3
      =   - \frac{k^2 N^3-3kN^2}{2}.
            \label{anomaliestrinion}
     \eea
  Note that these anomalies are independent of the $\fsu(k) \times \fsu(k)$ fluxes. 
  For example when $N=k=2$, these are
    \bea
    \Tr R'
      =   - 10, ~~~
     \Tr t
      =     8, ~~~
     \Tr R'^3
      =      2, ~~~
     \Tr R'^2 t
      =      4, ~~~
     \Tr R' t^2
      =    -4, ~~~
     \Tr t^3
      =   - 4.
    \eea
  Interacting trinions for this case were constructed and studied in \cite{Razamat:2016dpl}. Particularly there are three theories appearing there that have this $\fu(1)_s$ flux, with different $\fsu(k) \times \fsu(k)$ fluxes. These were dubbed $T_A$, $T_B$ and the $\frak{so}(5)$ trinion that has a Lagrangian description. In all three cases the anomalies agree with the result above.
   
  With these anomalies one can obtain those of the trinion with arbitrarily flux $N_s$ associated to the surface, say $N_s = \frac{n}{2}$,
  which are given by
    \bea
     \Tr R'
     &=&  - \frac{(k^2-2)(N-1) + 3k(N^2-1)}{2}, ~~~~
     \Tr t
      =   - \frac{k^2N n - 3 k N^2}{2} ,
             \nonumber \\
     \Tr R'^3
     &=&     \frac{(N-1)(k^2(N^2+N-1)+2) - 3 k (N^2-1)}{2}, 
               ~~~~~
     \Tr R'^2 t
      =     \frac{k^2 N (N^2 - 1)}{6} n,
               \nonumber \\
     \Tr R' t^2
     &=&   - \frac{k^2 N (N^2-1)}{6}, ~~~~~
     \Tr t^3
      =   - \frac{k^2 N^3 n-3kN^2}{2}.
			\label{anomaliestriniongen}
    \eea
   
  In the same way, the anomalies of the ``conjugate" trinon with $N_S = -\half$ and the three punctures with $\sigma=-1$ can be easily obtained. 
  The answer is simply \eqref{anomaliestrinion} where the signs of the anomalies involving odd power of $t$ are changed.

We can now subject this construction to the following consistency conditions. Take $2(g-1)$ trinions, each with some value of the $\fu(1)_s$ flux $n_i$, and glue them together. In this manner we get a genus $g$ Riemann surface with flux $\sum_i n_i$. Consistency now demands that taking $2(g-1)$ trinions and $3(g-1)$ $\Phi$ gluings we should recover (\ref{genusg}) with $N_s=\sum_i n_i$. It is straightforward to show that this is indeed true.

We can further complicate by adding the conjugate trinion. Consider taking $2g-2 - a$ trinions with fluxes $n_i$ and $a$ conjugate trinions with fluxes $n_j$ to build a genus $g$ Riemann surface with flux $\sum_i n_i - \sum_j n_j$. Punctures of opposite sign are glued together by $S$ gluing.
 Now when constructing the Riemann surface we use some combination of $\Phi$ gluing of punctures with a positive sign, which contribute the anomalies in \eqref{anomaliestube}, $\Phi$ gluing of punctures with a negative sign, which contribute the anomalies in \eqref{anomaliestube} but with $t \rightarrow - t$, and $S$ gluing which only contributes to the $R$ symmetry anomalies where it gives the same contribution as in \eqref{anomaliestube}. In fact we can construct the same theory in different ways using different combinations of the above. Particularly say we use $b$ $\Phi$ gluings of punctures with a negative sign then we must use $3a-2b$ $S$ gluings and $3(g-a-1)+b$ $\Phi$ gluings of punctures with a positive sign. It is straightforward to show that due to the structure of the contributions $b$ will drop out as required and further that summing all contribution we indeed recover \eqref{genusg}. 

It is straightforward to generalize this to more complicated cases. First we can consider trinions with punctures of different signs. We can also consider anomalies involving the $\fu(1)_{\beta_i}$ and $\fu(1)_{\gamma_i}$ symmetries. These will be sensitive also to the fluxes under these symmetries, and to the colors of the punctures. These generalizations should be messy, but straightforward and we shall not carry them out here. 

Let us here mention a caveat of the construction. We try to predict the anomalies of certain trinions under the assumption that these trinion have certain symmetries. It might happen that the puncture symmetries in certain situations are inconsistent with preserving some of the $\fu(1)_\beta^{k-1}\times \fu(1)_t\times\fu(1)^{k-1}_\gamma$ symmetries. Evidence for this was found in \cite{Razamat:2016dpl} where certain trinions were possible to construct only under assumptions that some of the symmetries are broken. This should be related to issues with discrete fluxes we have studied in the previous section and we do not study this question in the current context here.

\section{Fluxes for $\fu(1)_t$ symmetry for $k,N=2$}
\label{sec:appendix}

In the case of two M5 branes and ${\mathbb Z}_2$ singularity we can construct field theories associated to non vanishing flux of $\fu(1)_t$. These theories do not have a regular Lagrangian description, rather are described by Lagrangians with parameters fine tuned in strong coupling domain. To construct a general model we first build a theory
corresponding to sphere with two maximal punctures of same color and having a flux corresponding to $\fu(1)_t$.  The construction is based on singular Lagrangians one can obtain for models in this class derived in \cite{Razamat:2016dpl}  following \cite{Gadde:2015xta}.
To obtain such a tube model  we start with the $T_A$ trinion of \cite{Razamat:2016dpl} which has fluxes $(\frac14,\frac14,1)$ for $(\fu(1)_\gamma,\fu(1)_\beta,\fu(1)_t)$ associated to the surface. Then we flip the sign of one of the punctures by flipping the mesons corresponding to it \cite{Gaiotto:2015usa} (see \cite{Gaiotto:2012xa,Beem:2012yn,Agarwal:2015vla,Fazzi:2016eec,Nardoni:2016ffl} for similar procedure when $k=1$). The procedure of flipping the sign does not change the fluxes. Then we close the flipped maximal puncture first to minimal with vacuum expectation value for a meson shifting fluxes with $(-\frac14,\frac14,\frac12)$. Finally we close the minimal puncture  with vacuum expectation value for a baryon shifting fluxes with $(0,-\frac12,\frac12)$. The theory in the end has two maximal punctures of the same color and flux $(0,0,2)$. We can then insert this theory in our construction of torus models together with free trinions and close the minimal punctures to obtain theories with flux for all three symmetries. 

We now construct the tube model in more detail. We refer the interested readers for details to \cite{Razamat:2016dpl} and here we just summarize the construction of the $T_A$ trinion. The $T_A$ trinion can be built by taking a sphere with two minimal and two maximal punctures and flux $(0,0,1)$, which is constructed by combining together two free trinions,  and tuning to the point on the conformal manifold of the model where the abelian symmetries coming from minimal punctures enhance to $\fsu(2)\times \fu(1)_c$ and gauging the $\fsu(2)$ with special choice of matter. With that choice of matter the $\fu(1)_c$ symmetry enhances also to $\fsu(2)$ and one obtains additiona $\fsu(2)$ factor rotating the additional matter fields. We thus obtain model with three factors of $\fsu(2)\times \fsu(2)$ symmetry associated to a triplet of maximal punctures.  We can thus begin our procedure of flipping and closing maximal puncture by closing the puncture of the sphere with two minimal and two maximal punctures to obtain a sphere with two minimal and one maximal punctures and then perform the gauging needed to obtain the interacting trinion.

First we  flip the sign of one of the maximal punctures. This is done by flipping the mesons associated to that puncture \cite{Gaiotto:2015usa}. We add singlet fields $m_i$ and couple them through a superpotential to the mesons, $W=m\cdot M$. Next we close the puncture by giving vacuum expectation values to a particular combinations of $m$. As the mesons in the sphere with two minimal and two maximal punctures are built from $Q\widetilde Q$ combinations of chiral fields, the vacuum expectation values induce mass terms  for some of the flavors. The sphere with one maximal and two minimal punctures we obtain thus has Lagrangian in terms of two $\fsu(2)$ gauge groups each having five flavors and a bunch of singlets. 

Let us figure this out in complete detail.
The discussion is most easily performed at the level of the index as it captures all the relevant information.  The index of the sphere with two maximal and two minimal punctures is

\be\label{sphtt}
&&{\cal I}_{{\bf z},{\bf v}, a,b} =(p;p)^2(q;q)^2\oint\frac{dw_1}{4\pi i w_1}\oint\frac{dw_2}{4\pi i w_2}
\frac{\Gamma_e(\frac{pq}t (\beta\gamma)^{\pm1}w_1^{\pm1}w_2^{\pm1})}{\Gamma_e(w_1^{\pm2})\Gamma_e(w_2^{\pm2})}\\
&&\qquad \Gamma_e(t^{\frac12}\beta b^{-1}w_1^{\pm1}z_1^{\pm1})\Gamma_e(t^{\frac12}\gamma^{-1} b w_1^{\pm1}z_2^{\pm1})\Gamma_e(t^{\frac12}\gamma b w_2^{\pm1}z_1^{\pm1})\Gamma_e(t^{\frac12}\beta^{-1} b^{-1}w_2^{\pm1}z_2^{\pm1})\nonumber\\
&&\qquad \Gamma_e(t^{\frac12}\gamma a w_1^{\pm1}v_1^{\pm1})\Gamma_e(t^{\frac12}\beta^{-1} a^{-1}	 w_1^{\pm1}v_2^{\pm1})\Gamma_e(t^{\frac12}\beta a^{-1}w_2^{\pm1}v_1^{\pm1})\Gamma_e(t^{\frac12}\gamma^{-1} a w_2^{\pm1}v_2^{\pm1})\,.\nonumber
\ee We flip the ${\bf z}$ puncture,

\be
{\cal I}_{{\bf z},{\bf v}, a,b} \qquad\;\, \to \;\,\qquad \Gamma_e(\frac{p q}t (\gamma/\beta)^{\pm1}z_1^{\pm1}z_2^{\pm1}){\cal I}_{{\bf z},{\bf v}, a,b} \,.
\ee Then we give vacuum expectation values for the meson weighed $\frac{p q}t \frac\beta\gamma (z_1z_2)^{-1}$. In the index we compute residue when $z_1=(\frac{pq}t)^{\frac12}\beta\epsilon$, $z_2=(\frac{p q}t)^{\frac12}\gamma^{-1}\epsilon^{-1}$. We also need to introduce new chiral fields coupling them through superpotential~\cite{Gaiotto:2015usa}. In the index this amounts to multiplying with
$\Gamma_e(pq\beta^{2}\gamma^{-2})\frac{\Gamma_e(\frac{p q}t\beta^{2}\epsilon^{-2})\Gamma_e(\frac{p q}t\gamma^{-2}\epsilon^2)}{\Gamma_e(\frac{p q}t\beta^{2}\epsilon^2)\Gamma_e(\frac{p q}t\gamma^{-2}\epsilon^{-2})}$. Then we need to give a vacuum expectation value to a specific baryonic operator which amounts to setting $\epsilon=(\frac{t}{p q})^{\frac12}\gamma^{-1}$. We also need to flip the second baryon charged under $u(1)_\gamma$ by multiplying the index with $\Gamma_e(pq\gamma^4)$. After all the above manipulations index of the sphere with two minimal and one maximal puncture we obtain is,

\be\label{infh}
{\cal I}_{{\bf v},a,b}= \Gamma_e(\frac{(p q)^2}{t^2})\Gamma_e(\frac{(p q)^2}{t^2}(\frac{\gamma}{\beta})^{\pm2})\Gamma_e(\frac{(p q)^2}{t^2}(\gamma\beta)^2)\, {\cal I}_{\{z_1=\frac\beta\gamma,z_2=\frac{p q}t\},{\bf v}, a,b}\,,\ee Let us look at ${\cal I}_{\{z_1=\frac\beta\gamma,z_2=\frac{p q}t\},{\bf v}, a,b}$. Two of the fundamental fields in the second line of \eqref{sphtt} form mass term and decouple. Thus the index becomes,

\be
&&{\cal I}_{\{z_1=\frac\beta\gamma,z_2=\frac{p q}t\},{\bf v}, a,b} = (p;p)^2(q;q)^2\oint\frac{dw_1}{4\pi i w_1}\oint\frac{dw_2}{4\pi i w_2}
\frac{\Gamma_e(\frac{pq}t (\beta\gamma)^{\pm1}w_1^{\pm1}w_2^{\pm1})}{\Gamma_e(w_1^{\pm2})\Gamma_e(w_2^{\pm2})}\\
&&\;\;\;\,\qquad \Gamma_e(t^{\frac12}\frac{\beta^2}\gamma b^{-1}w_1^{\pm1})\Gamma_e(\frac{t^{\frac32}}{   p   q  }\gamma^{-1} b w_1^{\pm1})\Gamma_e(t^{\frac12}\frac{\gamma^2}\beta b w_2^{\pm1})\Gamma_e(\frac{t^{\frac32}}{  p  q  }\beta^{-1} b^{-1}w_2^{\pm1})\nonumber\\
&&\qquad \Gamma_e(t^{\frac12}\gamma a w_1^{\pm1}v_1^{\pm1})\Gamma_e(t^{\frac12}\beta^{-1} a^{-1}	 w_1^{\pm1}v_2^{\pm1})\Gamma_e(t^{\frac12}\beta a^{-1}w_2^{\pm1}v_1^{\pm1})\Gamma_e(t^{\frac12}\gamma^{-1} a w_2^{\pm1}v_2^{\pm1})\,.\nonumber
\ee Which is the index of two copies of  $\fsu(2)$ SQCD with five flavors coupled together through superpotential terms and bifundamental fields.

Finally the index of the tube theory with flux for $\fu(1)_t$ and two maximal punctures of the same color is obtained by gauging the $\fsu(2)$ enhances group of $\fu(1)_a$ and $\fu(1)_b$ with appropriate chiral fields,

\be
&&{\cal I}_{{\bf v},\bf {c}} =\Gamma_e(t (\frac\gamma\beta  v_2)^{\pm1}v_1^{\pm1})
\Gamma_e(p\,q \frac{1}{\beta^2\gamma^2})
\\
&&\;\,\; (p;p)(q;q)\oint\frac{dz}{4\pi i z}\frac{\Gamma_e(\frac{pq}{t \gamma\beta}(\beta\gamma^{-1} v_2^{-1})^{\pm1} z^{\pm1})}{\Gamma_e(z^{\pm2})}\Gamma_e(\gamma\beta z^{\pm1}v_1^{\pm1}){\cal I}_{{\bf c}, \sqrt{z v_2},\sqrt{v_2/z}}\,.\nonumber
\ee We have ${\cal I}_{{\bf c}, \sqrt{z v_2},\sqrt{v_2/z}}$ written in \eqref{infh}. We note that setting $\beta\gamma$ to be one the above index smoothly becomes

\be
&&{\cal I}_{{\bf v},\bf {c}}  \to \Gamma_e(\frac{(p q)^2}{t^2})^2\Gamma_e(\frac{(p q)^2}{t^2}\beta^{\pm4}){\cal I}_{\{z_1\to\frac\beta\gamma,z_2\to\frac{p q}t\},{\bf c}, \sqrt{z v_2},\sqrt{v_2/z}}\left.\right|_{\beta\gamma\to 1}\,.\nonumber
\ee Smooth deformation of this sort breaking symmetry might correspond to marginal deformations. Indeed we do expect to have marginal deformations of this sort which break puncture symmetries down to the Cartan~\cite{Razamat:2016dpl}. We thus have evidence that the tube theory corresponds to sphere with four minimal punctures and flux two for $\fu(1)_t$ when we tune the couplings to point where the abelian symmetries enhance.

The anomalies of this theory are easily computed from Lagrangian implied by this index. 
Let us write the trial a anomaly for the model one obtains gluing together ${\frak Q}^\beta+{\frak Q}^\gamma$ free trinions and ${\frak Q}_t$ tubes with $\fu(1)_t$ flux. We then clos minimal punctures to obtain ${\frak Q}_\beta/2$ units of $\beta$ flux and  $-{\frak Q}^\gamma/2$ units of $\gamma$ flux,  and $2{\frak Q}_t$ units of $t$ flux. The anomalies are given, 

\be
&&a = \frac{3}{4} ({\frak Q}^\beta {s_\beta} (15 {s_\beta}^2+9 {s_\gamma}^2+9 {s_t}^2-5)-2 {\frak Q}_t {s_t} (18 {s_\beta}^2+18 {s_\gamma}^2+6 {s_t}^2-5)+\\&&\;\,\;\,\;\quad{s_\gamma} {\frak Q}^\gamma (9 {s_\beta}^2+15 {s_\gamma}^2+9 {s_t}^2-5))\,,\nonumber
\ee from which we determine the cubic anomalies which depend on flux of $\fu(1)_t$,

\be
k_{t\beta\beta} = -32\,\; {\frak Q}_t\,\qquad k_{t\gamma\gamma} = -32\,\; {\frak Q}_t\,, \qquad k_{ttt} = -32\,\; {\frak Q}_t\,.
\ee We can compute the dependence of $TrR$ on ${\frak Q}_t$ and get that it is equal to $-8s_t{\frak Q}_t$ which tells us that $\Tr \fu(1)_t=-8{\frak Q}_t$. All other anomalies do not depend on ${\frak Q}_t$. The anomalies here coincide with the ones deduced from  \eqref{eighf}.

\bibliographystyle{ytphys}
\bibliography{refs}

\end{document}